\documentclass[aps,showpacs]{revtex4}

\usepackage{psfig}
\usepackage{epsfig}
\usepackage{amssymb}
\usepackage{amsmath}

\begin{document}

\title{ The random K-satisfiability problem: from an analytic solution
to an efficient algorithm}

\author{ Marc M\'ezard$^{1}$ and Riccardo Zecchina$^2$ }

%\author{Marc M\'ezard} \email{mezard@ipno.in2p3.fr}
%\affiliation{Laboratoire de Physique Th\'eorique et Mod\`eles
%Statistiques, Universit\'e Paris Sud, 91405 Orsay, France}
%
%\author{Riccardo Zecchina} \email{zecchina@ictp.trieste.it}
%\affiliation{International Center for Theoretical Physics, Statistical
%Mechanics and Interdisciplinary Applications Group, Strada Costiera
%11, I-34100 Trieste, Italy}

\affiliation{$^{1}$Laboratoire de Physique Th\'eorique et Mod\`eles
Statistiques, CNRS and Universit\'e Paris Sud, B\^at. 100, 91405 Orsay
{\sc cedex}, France\\ $^2$The Abdus Salam International Centre for
Theoretical Physics,\\ Statistical Mechanics and Interdisciplinary
Applications Group,\\ Str. Costiera 11, 34100 Trieste, Italy }

\date{\today}

%\maketitle

\begin{abstract}
We study the problem of satisfiability of randomly chosen clauses,
each with K Boolean variables.  Using the cavity method at zero
temperature, we find the phase diagram for the K=3 case. We show the
existence of an intermediate phase in the satisfiable region, where
the proliferation of metastable states is at the origin of the
slowdown of search algorithms.  The fundamental order parameter
introduced in the cavity method, which consists of surveys of local
magnetic fields in the various possible states of the system, can be
computed for one given sample. These surveys can be used to invent new
types of algorithms for solving hard combinatorial optimizations
problems.  One such algorithm is shown here for the 3-sat problem,
with very good performances.
\end{abstract}


\maketitle
%\begin{multicols}{2}
%\narrowtext

\section{Introduction}

The K-satisfiability (Ksat) problem deals with an ensemble of $N$
Boolean variables, submitted to $M$ constraints.  Each constraint is
in the form of an 'OR' function of $K$ variables in the ensemble (or
their negations), and the problem is to know whether there exists one
configuration of the variables (among the $2^N$ possible ones) which
satisfies all constraints.  The Ksat problem for $K\geq 3$ is a
central problem in combinatorial optimization: it was the first
problem to be shown NP-complete~\cite{Cook,Garey_Johnson}, and an
efficient algorithm for solving Ksat in its worst case instances would
immediately lead to other algorithms for solving efficiently thousands
of different hard combinatorial problems.

At the core of the statistical physics of disordered systems is the
spin glass problem (SG), which also deals with Boolean variables
('spins'), interacting with random exchange couplings \cite{MPV}.
Each pair of interacting spins can be seen as a constraint, and
finding the state of minimal energy in a spin glass amounts to
minimizing the number of violated constraints. Although the precise
form of the constraints in SG and Ksat differ, there exist deep
similarities\cite{MoZe_prl,MoZe_pre}; in both cases the difficulty
comes from the existence of 'frustration'\cite{MPV}, which forbids to
find the global optimal state by a purely local optimization
procedure.  Links between combinatorial optimization and statistical
physics have been known for long\cite{MPV}. Two main categories of
questions can be addressed. One type is algorithmic, for instance
finding an algorithm which decides whether an instance is SAT or
UNSAT. Another is more theoretical, and deals with large random
instances, for which one wants to predict the typical
behaviour. Examples of use of statistical physics in each category are
the simulated annealing algorithm \cite{SimAnn} and the solution of
the random assignment problem \cite{Match,Aldous}, or the direct
mapping of certain graph partitioning problems to spin glasses
\cite{FuAnd}. Here we address the two types of questions in the 3sat
problem.

The study of random Ksat problems, where the clauses are chosen
randomly, is also interesting from the viewpoint of optimization. In
practice, algorithms which are used to solve real-world NP-complete
problems display a huge variability of running times, ranging from
linear to exponential, when the parameters (e.g. the number of
clauses) are changed. A theory for the typical-case behaviour of
algorithms, on classes of random instances chosen from a given
probability distribution, is therefore the natural complement to the
worst-case
analysis~\cite{average1,average2,average3,AI_issue,Cook_review,TCS_issue}.

In random 3sat, numerical simulations have shown the existence of a
phase transition when one varies the ratio $\alpha=M/N$ of the number
of clauses to the number of variables. For $\alpha<\alpha_c$ the
generic problem is satisfiable (SAT), for $\alpha>\alpha_c$ the
generic problem is not satisfiable (UNSAT)\cite{KirkSel}.  Using the
cavity method, first developed in spin glass theory, we shall show the
existence of this threshold and compute $\alpha_c\simeq 4.267$.  We
also find an intermediate region, the HARD-SAT-Phase
$\alpha_d=3.921<\alpha <\alpha_c$, where the generic problem is still
SAT but the proliferation of metastable states makes it difficult
for algorithms to find a solution. This proliferation is similar to
the effect found in the theories of structural glasses using spin
glass models with multispin interactions \cite{pspin}, where it is
known to lead to a dramatic slowdown of the relaxation.  In this sense
the difficulty to solve the 3sat problem in the intermediate region
$\alpha_d<\alpha <\alpha_c$ is similar to the difficulty in
equilibrating structural glasses.

This theoretical analysis is done using the cavity method, at a level
equivalent to what is called one step replica symmetry breaking in the
replica language. This means that it assumes the existence of many
states, but cannot handle a nontrivial correlation pattern between
them. There are some arguments which point towards the correcteness of
this solution, although an exact proof looks somewhat remote at
present.

In this cavity method with many states, the order parameter consists
in the surveys of local magnetic fields acting on each spin.  While
for the theoretical analysis one averages over the random graph
structure of the problem, it turns out that this order parameter can
also be computed for one given sample, using a reasonably simple
message passing procedure which takes into account the multiplicity of
states.  This procedure provides a generalization of the belief
propagation used in statistical inference \cite{Yedidia}; it is shown
here to converge in some difficult situation with many equilibrium
states where ordinary belief propagation does not converge. The
resulting surveys provide an interesting description of one given
sample, where the various variables are found to play very different
roles. This single sample analysis is very useful in order to find new
algorithms for solving hard optimization problems.  Here we show one
such algorithm for the random 3sat problem, where the surveys are used
to identify one spin and fix it. The problem is thus reduced and the
surveys are computed again on the new system.  This decimation
procedure is shown to have very good performances, comparable to, or
better than, the state of the art in this problem.

The paper presents a number of concepts and techniques, both
analytical and numerical, which can be applied to a rather large class
of combinatorial optimization problems.  We have presented these
concepts and techniques in a broad framework, in order to allow for
future use on different problems. The concrete implementation is then
done on the random 3sat problem. Some of the results discussed in this
paper have been recently announced in \cite{MEPAZE}.

The paper is organised as follows: in sect. \ref{sect_fact_graph} we
present the generic structure of the optimization problems in which we
are interested.  These can be represented as bipartite graphs called
factor graphs. Sect.  \ref{sect_sum_prod} recalls a general message
passing procedure which can be used to study optimization or inference
problems defined on these factor graphs.  The basic ingredients of
this procedure are messages which we call cavity-biases which play a
crucial role in the whole paper. Sect. \ref{sect_rangraph} defines the
set of random graphs which we study, which are random hypergraphs with
a fixed connectivity. Sect. \ref{sect_states} provides some background
on the decomposition of the configuration space into states. It
consists of a short introduction for non-physicists, a specific
definition of zero temperature states for the random 3sat problem, and
the definition of the {\sl complexity} which is a crucial concept for a
system with many states.  Sect. \ref{sect_cavity} provides an
introduction to the zero temperature cavity method, presented in the
general setting of combinatorial problem on random factor graphs. This
section summarizes, and puts in a more general context, some recent
work which has developed the cavity method for finite connectivity
problems, first at finite temperature \cite{MP_Bethe}, then at zero
temperature \cite{MP_T0}. It provides the whole formalism for the
analytic study of the phase diagram.  This formalism is applied to the
random 3sat problem in sect. \ref{sect_r3sat}, where all the results
on the phase diagram are derived. We explain the survey propagation
algorithm on a given sample in sect. \ref{sect_SP}, and the decimation
algorithm for solving large random 3sat problems is presented in
\ref{sect_SID}.  Sect. \ref{sect_concl} contains some concluding
remarks.  Appendix A contains some technical details relative to the
computation of the phase diagram done in sect.\ref{sect_r3sat}.
Appendix B explains the computation of the free energy for one given
sample.

\section{Factor graph representation }
\label{sect_fact_graph}
The models we are interested in involve Boolean variables which
interact in groups, the energy being the sum of energies of all
groups.  We shall adopt the factor graph
representation~\cite{factor_graph} familiar in computer science, but
we shall keep to the representation of Boolean variables as Ising
spins, more familiar to statistical physicists.

We consider a set of $N$ Ising spins, $\sigma_i \in\{ \pm 1 \}$, and
we suppose that we have $M$ groups of interacting variables, which are
called function nodes. Each function node $a$ involves a set of $n_a$
spins.  We denote by $V_a$ the set of all these spins. The interaction
is an arbitrary function of the spins in $V_a$, which depends on the
problem one considers, and can also involve hidden variables.

The total energy of a configuration $\sigma_1,...,\sigma_N$ is 
\begin{equation}
E=\sum_{a=1}^M E_a \ ,
\end{equation}
and the goal in combinatorial optimization is to find a configuration
of spins which minimizes $E$.  A generalization of this problem,
natural from the point of view of physics, and which connects with
problems in statistical inference, consists in introducing an
additional parameter $\beta$, an 'inverse temperature' in the physics
language, and in studying the Boltzmann probability distribution:
\begin{equation}
P(\sigma_1,...,\sigma_N)=\frac{1}{Z} \exp\left( -\beta E\right) \ ,
\label{boltz}
\end{equation}
where $Z$ is a normalisation constant. As usual in physics, we shall
denote by $<O>$ the expectation of an observable $O$ (which can be any
function of the $\sigma_i$) with respect to this measure.  In the
large $\beta$ ('low temperature') limit this measure concentrates onto
the lowest energy configurations. At finite $\beta$ one may be
interested in computing for instance the expectation value of one spin
variable $\sigma_1$ with the Boltzmann probability.  Because we work
with binary spins, this average determines the full marginal
probability law of $\sigma_1$: this is precisely the quantity that one
typically seeks in many inference problems, like e.g. decoding
procedures for error correcting codes.

\begin{figure}
\centering
\includegraphics[width=8.cm]{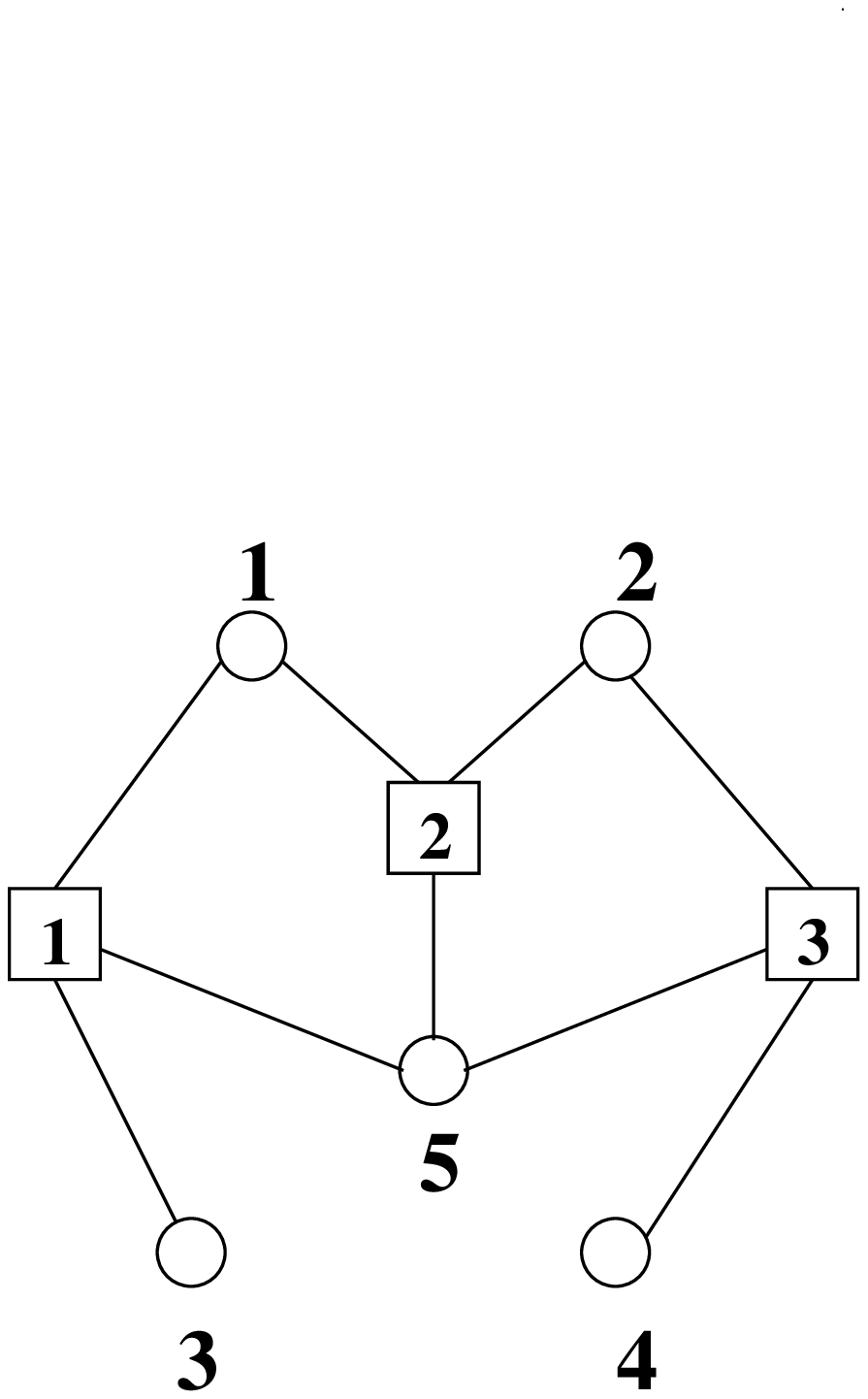}
\caption{
An example of a factor graph with $5$ variable nodes $i=1,..,5$ and
$3$ function nodes $a=1,2,3$. In this case, each function node has connectivity
$3$, as in the 3sat problem. The connectivities of the $5$  variable nodes are 
respectively $2,2,1,1,3$.}
\label{fig_graph}
\end{figure}

The general problem can be represented by a graph consisting of two
types of vertices, 'variable nodes' associated with each spin, and
'function nodes'. A function node $a$ is connected by edges to all the
variable nodes involved in $S_a$.  Therefore each variable node has
connections towards all the function nodes in which it appears, and
the graph is bipartite (see fig.\ref{fig_graph}). Each spin $\sigma_i$
is connected to $n_i$ function nodes, we denote this set of function
nodes by $V_i$.  We call $n_i$ the connectivity of spin $i$, $n_a$ the
connectivity of function node $a$.  Throughout this paper, the
variable nodes indices are taken in $i,j,k,...$, while the function
nodes indices are taken in $a,b,c,...$

Let us give here a few standard examples.

Spin glasses: all the interactions involve two spins, so all $n_a$ are
equal to $2$; the energy of an interaction node $a$ involving spins
$\sigma_i$ and $\sigma_j$ is given by
\begin{equation}
E_a=-J_{ij} \sigma_i \sigma_j \ ,
\end{equation}
 where the number $J_{ij} $ is called
the coupling constant. Generalized spin glasses with p-spin interactions,
\begin{equation}
E_a=-J_{{i_1}...{i_p}} \sigma_{i_1}...\sigma_{i_p} \ ,
\label{epsdefpspin}
\end{equation}
have also been  studied a lot in statistical physics as  models of structural 
glasses. They are the closest physical analogues of the
satisfiability problems which we study here \cite{SimpleKsat}.

K-sat: All interactions involve  $K$ spins, and the energy
of an interaction node $a$ involving spins $\sigma_{i_1},...,
\sigma_{i_K}$ is given by:
\begin{equation}
E_a= 2 \prod_{r=1}^K \frac{\left(1+J_a^r \sigma_{i_r}\right)
}{2} \ .
\label{epsdefKsat}
\end{equation}
It depends on a set of $K$ coupling constants ${\bf
J_a}=(J_a^1,...,J_a^K)$ which take values $\pm 1$. This interaction
node has a simple interpretation as a clause: the energy $E_a$ is zero
as soon as at least one of the spins $\sigma_{i_r}$ is opposite to the
corresponding coupling $J_a^r$.  If all spins are equal to their
couplings, the energy is equal to $2$.  The more conventional
description of Ksat uses Boolean variables: let us introduce $x_i$
which is TRUE if and only if $\sigma_i=1$.  The energy $E_a$ depends
on the OR of the $K$ variables $y_{i_1}$,...$y_{i_K}$, where $y_{i_r}$
is the original $x_{i_r}$ when $J_a^r=-1$ and is its negation when
$J_a^r=+1$. The energy vanishes if $y_{i_1} \vee ... \vee y_{i_K}$ is
true (the clause is then said to be satisfied), otherwise it is equal
to $2$ and the clause is unsatisfied. This arbitrary factor of $2$ is
introduced for future convenience.

One can also consider graphs involving mixtures of function nodes of
different types, e.g. mixtures of $K=2$ clauses and $K=3$ clauses
\cite{MoZe_pre,MZKST}.  These are some examples of constraints
satisfaction problems, but of course there exist many other instances
of problems, much studied in computer science, which can be
represented by such factor graphs.

In general, an instance of the problem (also called a sample in
physics language) is given by a graph and the set of couplings needed
to define each function node. In physics (e.g. spin glasses) one is
interested in the configurations of low energy. In the SAT problem,
one wants to know whether there exists a configuration of zero energy
(in which case the instance is called SAT), or not (in which case the
instance is UNSAT).

\section{The sum-product algorithm}
\label{sect_sum_prod}
A popular method for studying the inference, i.e. the probability
measure (\ref{boltz}), is a message passing procedure called the
``Sum-Product'' algorithm\cite{factor_graph,Yedidia}. When used at
$\beta \to \infty$, the corresponding ``Min-Sum'' algorithm can also
be used to get some information on the lowest energy configurations.
This procedure is exact and fast on tree-like graphs.  In our case the
Sum-Product algorithm amounts to sending some messages along the edges
of the graph. We call cavity-field, and denote by $h_{i \to a}$, the
message passed from a variable node $i$ to a function node $a$. We
call cavity-bias, and denote by $u_{a \to i}$, the message passed from
a function node $a$ to a variable node $i$.

The cavity-field $h_{i \to a}$ is given by the sum of cavity-biases
converging to $i$ from all function nodes $b$ distinct from $a$:
\begin{equation}
h_{i \to a} = \sum_{b \in V(i)/a} u_{b \to i}
\label{sum_u}
\end{equation}

The operation performed by a function node to compute the
cavity-biases which it will send to its neighboring variable nodes is
a partial summation: it computes the marginal probability law for that
variable to which it sends the message. More precisely, let us
consider a function node $a$ of connectivity $K$ and let us suppose,
for notational simplicity, that it is connected to variables
$\sigma_1,...,\sigma_K$. The cavity-bias $u_{a \to 1}$ sent from the
function node $a$ to the variable node $i=1$ is a function of the
cavity-fields $h_{j\to a}$ sent from all other variables nodes $j \in
\{ 2,...,K \}$ towards node $a$.  One considers the function of
$\sigma_1$defined by $ \sum_{\sigma_2,...,\sigma_K} \exp\left(-\beta
E_a(\sigma_1,\sigma_2,...,\sigma_K) +\beta \left(h_{2 \to a}
\sigma_2+...+ h_{K \to a} \sigma_K\right)\right) $. As $\sigma_1 = \pm
1$, this function can be written for instance as the exponential of a
linear form in $\sigma_1$. The cavity-bias $ u_{a \to 1}$ sent from
$a$ to $1$ is defined from:
\begin{equation}
%\frac{
\sum_{\sigma_2,...,\sigma_K} 
\exp
\left(
-\beta 
E_a(\sigma_1,\sigma_2,...,\sigma_K) +\beta 
		\left( h_{2 \to a} \sigma_2+...+
						h_{K \to a} \sigma_K
    \right)
\right)
= \exp\left(\beta \left(w_{a \to 1}+\sigma_{ 1} u_{a \to 1}\right)
\right) \ .
\label{uwdef}
\end{equation}
Beside the cavity-bias $ u_{a \to 1}$, this equation also defines a
'free-energy shift' $w_{a \to 1}$ which is not used in the Sum-Product
algorithm but will become very important in our generalisation later
on.  In physics words, $h_{i \to a}$ is the magnetic field on spin
number $i$ whenever the interaction $a$ is turned off, and $ u_{a \to
i}$ is the contribution to the magnetic field on spin number $i$ from
the interaction $a$. Equation (\ref{sum_u}) indicates that the
probability law of spin $\sigma_i$ due to the interactions $a=2,..K$ is a
product of independent laws due to each interaction, while the
marginalization operation (\ref{uwdef}) is a partial summation, hence
the name sum-product.  The algorithm is easily generalized to
variables which are more complicated than Boolean.

The iteration of the above message passing algorithm, starting from a
generic random initial condition, is known to converge whenever the
underlying factor graph is a tree. Actually it converges in one sweep
if one first computes the messages from the leaves of the tree.  The
resulting set of messages can be used to compute the probability
distribution of one spin (or more generally of some subset of spins).
One just needs to compute the local field $H_i$ on spin $\sigma_i$:
\begin{equation}
H_i=\sum_{a \in V(i)} u_{a \to i} \ ,
\label{Hdef}
\end{equation}
and the probability distribution of $\sigma_i$ is:
\begin{equation}
P(\sigma_i)=\frac{\exp\left( \beta H_i\sigma_i\right) }
{2 \cosh \left( \beta H_i \right)}\ .
\label{local_meas}
\end{equation}
One way to prove this result, using the physics language, is to show
that the message passing algorithm minimizes the Bethe free energy of
the spin system \cite{Yedidia}. As the Bethe free energy is exact for
tree like graphs, this provides the proof.  If one is interested in
the optimization problem ($\beta \to\infty$), one can show that the
configuration $\sigma_i=\mbox{Sign} H_i$ is the lowest energy
configuration if there is a unique such configuration.  If there are
several lowest energy configurations, taking the $\beta \to \infty$
limit of $<\sigma_i>$ with the measure (\ref{local_meas}) gives the
average of the spin $\sigma_i$ over all these configurations (but one
needs to make this detour through the finite $\beta$ problem in order
to get this result).

In this work we shall mainly be interested in the optimization
problem.  Working directly at $\beta=\infty$ then simplifies the
algorithm.  The cavity-fields are generated as before by sums of
cavity-biases.  For computing a cavity-bias, one performs a partial
minimization, and the formula (\ref{uwdef}) simplifies to
\begin{equation}
\min_{\sigma_2,...,\sigma_k} 
\left[
E_a\left(\sigma_1,\sigma_2,...,\sigma_k\right) - 
\left(h_{2 \to a} \sigma_2+...+
     h_{k \to a} \sigma_k
\right)
\right]
= - \left(w_{a \to 1}+\sigma_{ 1} u_{a \to 1}\right) \ .
\label{uwdef2}
\end{equation}
This equation defines the output messages $w$ and $u$ as functions 
of the input messages $h_{j \to a}$. In general, we shall write
\begin{equation}
w_{a \to 1}= \hat w_{\bf J_a}(h_{2 \to a},...,h_{K \to a}) \ \ ; \ \ 
u_{a \to 1}= \hat u_{\bf J_a}(h_{2 \to a},...,h_{K \to a}) \ ,
\label{hatdef}
\end{equation}
which defines the functions $\hat w_{\bf J_a}$ and $\hat u_{\bf J_a}$
(the label ${\bf J_a}$ is here to explicitly remind that a given
function node energy $E_a$ will in general depend on some set of
couplings - see (\ref{epsdefpspin},\ref{epsdefKsat})- which we denote
collectively as ${\bf J_a}$).

\section{Random graphs and  thermodynamic limit}
\label{sect_rangraph}
\subsection{Definition of random graphs}

In the rest of this paper we shall consider the random K-sat problem,
which is defined on some ensemble of random graphs which we now
describe.  To lighten the notation, we concentrate on the $K=3$ cases
for which all the function nodes have connectivity $n_a=3$, and we
generate the random graphs as follows: For each triplet $i<j<k$ of
variable nodes, a function node connecting them is present with a
probability $ 6 \alpha/N^2$, and it is absent with probability $1-6
\alpha/N^2$.  The average number of function nodes is then $M= \alpha
N$. The graph model used is
analogous to the G(N,p) model of random graph theory (see, e.g.
\cite{Bollobas}), with $p=6 \alpha/N^2$.

For the problem which we consider, the energy $E_a$ associated with a
function node $a$ also depends on some coupling constants ${\bf J_a}$
(see for instance (\ref{epsdefKsat})), which may be drawn randomly and
independently for each function node from some probability
distribution. For instance in random 3sat, each number
$J_a^1,J_a^2,J_a^3$ takes values $\pm 1$ with probability $1/2$.  In
general we shall denote by ${\cal E}_J O$ the average of any quantity
$O$ over all the choices of the random graphs (with fixed $N$ and
$M$), and over the choices of couplings.  In such a probabilistic
setting, one is interested for instance in computing the average
'ground state energy': For each sample, an optimal configuration (one
with minimal energy), is called the ground state, its energy is $E_0$,
and one would like to compute ${\cal E}_J E_0$.

\subsection{Thermodynamic limit}
We shall be interested in the '{\it thermodynamic limit}' where $M$
and $N$ go to $\infty$, keeping the ratio $\alpha=M/N$ fixed. The
connectivities of variable nodes become independent identically
distributed (iid) random variables with a Poisson distribution $f_{3
\alpha} (k)$ of mean $3 \alpha$, since the probability of having $k$
edges connected to a variable node is:
\begin{equation}
\lim_{N \to \infty} 
\left( 
\begin{array}{c}
(N-1)(N-2)/2 \\ k
\end{array} 
\right) 
(6 \alpha/N^2)^k (1-6 \alpha/N^2)^{(N-1)(N-2)/2-k}
 = \frac{(3 \alpha)^k}{k!} \exp(-3 \alpha) \equiv f_{3 \alpha} (k)
\label{Poisson}
\end{equation} 

The structure of the random graphs generated by this process for large
$N$ is interesting. Locally such a graph is tree-like: the typical
size of a loop in the graph scales like $\log(N)$ for large $N$. On
the other hand loops are definitely present, and they can induce
'frustration' in the sense of having competing constraints (it has
been argued that similar random graphs with a local tree-like
structure provide a natural setting for discussing the 'Bethe
approximation' of frustrated systems \cite{MP1}).  The structure of
the graph has one important consequence: consider one given function
node, connected to three variable nodes (spins). If one deletes this
function node, the typical distance between any two of these three
spins (measured as the length of the shortest path on the graph which
connects them) is of order $\log N$, and thus diverges in the
thermodynamic limit: the spins are far apart. This property will be
crucial in understanding the type of correlations existing between the
spins, and in solving the model.  Notice that the limit where $M,N \to
\infty$ is also the one that is interesting from a computational
complexity point of view.

For problems defined on random graphs with given $N,M$, the ground
state energy fluctuates from sample to sample.  It is often true, but
it may be difficult to show, that the distribution of the ground state
'energy density' $E_0/N$ becomes more and more peaked when $N$
increases, so that, in the thermodynamic limit, almost all samples
have the same energy density, which can be computed as
\begin{equation}
\epsilon_0 = \lim_{N \to \infty}{\cal E}_J E_0/N
\end{equation}
For random Ksat it can be proved that the above condition holds
\cite{Frieze}.

One of our aims is to compute this limiting value $\epsilon_0$ for a fixed
value of $\alpha=M/N$.  For the Ksat problem it turns out that this
value is equal to zero below a certain threshold $\alpha_c$, and
becomes $>0$ for $\alpha>\alpha_c$.  In statistical physics it is very
difficult to go beyond the estimate of the energy density: if one can
compute $\epsilon_0$, one knows that $ E_0 \sim \epsilon_0 N$ is the leading
behaviour of the energy for large $N$, but in general one cannot
control the subleading part, and in principle it could be possible for
instance that also in the small $\alpha$ phase of Ksat where
$\epsilon_0=0$, some finite contribution to $E_0$ (finite when $N \to
\infty$) could make the problem typically UNSAT.  Numerical
simulations tend to show that this is not the case.  Knowing $\epsilon_0$
will then allow to get the phase diagram of the problem. But one is
also interested in other properties of the generic samples in the
thermodynamic limit, like the decomposition of the space of accessible
configurations at a given energy, to which we now turn.

\section{States and clustering property}
\label{sect_states}
\subsection{A simple example of pure states: the ferromagnet}
One of the main aims of statistical physics is to understand the
building up of correlations between distant variables, when the basic
interactions between them are short range. This is precisely the type
of question that we need to address here: variables interact locally
(the only direct interactions involve spins connected to the same
function node).  But we also need to control the correlations
established between two spins belonging to the same function node, due
to their indirect coupling through other nodes. As we saw, the
geometry is such that this indirect interaction builds up through very
long ($O(\log N)$) paths.

Usually, in the statistical physics of systems with short range
interactions, the correlation between distant variables displays a
relatively simple variety of behaviours. The simplest one is when
there is only one pure state in the system (typically a 'paramagnetic
phase'): then there exists a finite correlation length and the
connected correlation function between two distant spins $\sigma_i,
\sigma_j$ decays exponentially with the distance $d_{ij}$ at large
distances:
\begin{equation}
\vert<\sigma_i \sigma_j>-<\sigma_i><\sigma_j> \vert
\simeq C \exp(-d_{ij}/\xi)
\end{equation}
where $\xi$ is the correlation length ($C$ can be a constant, or
involve power law corrections in the distance). This is called the
clustering property. On the other hand, some systems can also have
phase transitions, and display a low temperatue phase with several
pure state.

The archetypical case which we briefly describe here as a pedagogical
example is the ferromagnetic $p=2$ spin system with energy given by
(\ref{epsdefpspin}) with $J_{ij}=1$: at low temperature the spins
polarize in one of two 'pure states', related to each other by the
global symmetry changing all $\sigma_i$ to $-\sigma_i$.  Let us call a
configuration one assignment of the $N$ spins,
$\sigma_1,...,\sigma_N$.  The pure states are probability measures on
the configuration space obtained using a slightly modified Boltzmann
measure where one adds an external 'symmetry breaking' magnetic field
(in the present language one adds a function node of connectivity one
connected to each variable node $i$ , with energy $-B \sigma_i$). One
computes $\lim_{B \to 0^{\pm}} \lim_{N \to \infty} <\sigma_i>$, which
defines the expectation value $<\sigma_i>_{\pm}$ of spin $i$ in each
of the two states $+$ and $-$.  It is a well known fact that the
connected correlation functions within each state have the clustering
property:
\begin{equation}
\vert <\sigma_i \sigma_j>_\pm - <\sigma_i>_\pm <\sigma_j>_\pm \vert
\simeq C 
\exp(-d_{ij}/\xi)
\end{equation}
This means that when the spins collectively polarize in the $+$ state,
the correlations between distant spins vanishes. It is not true for
the full Boltzmann measure: if one does not add the small symmetry
breaking field $B$, but keeps to the Boltzmann measure (\ref{boltz}),
one gets for any observable $<O>=1/2 (<O>_+ + <O>_-)$ (the fact that
the two states enter with equal weight $1/2$ is a consequence of the
global symmetry of the original problem), and one easily shows that
the corresponding correlations do not vanish at large distances.

The lesson we learn from statistical physics is that correlations
decay at large distance within each pure state.  In problems more
complicated than the ferromagnet it may be difficult to identify the
various pure states, especially when we do not have at hand a simple
breaking of a symmetry. A large part of the work on spin glasses has
been devoted to this problem and we shall not try to reproduce it here
(see \cite{MPV} for a review), nor to give a general definition of
states at finite temperature in our problems.

\subsection{States at zero temperature}
Instead we shall focus on the zero temperature limit ($\beta \to
\infty$), where the situation is simpler. A state is defined in the
thermodynamic limit as a cluster of configurations, all of equal
energy, related to each other by single spin flip moves, and which are
'locally stable', in the sense that the energy cannot be decreased by
any flip of a finite number of spins. In the random 3sat problem one
can use an even simpler definition which allows to generalize the
definition of states also to finite $N$ problems: the condition of
local stability can be substituted by a condition of stability with
respect to any sequence of one spin flips.  The reason for this
simplification, specific to the Ksat problem, is that in this case the
stability with respect to sequences of single spin flips insure
stability with respect to collective flips of finite sets of spins.

\subsection{Many states: definition of the complexity}
Experience with disordered and frustrated systems like glasses shows
that there can exist many states, and the number of states typically
grows exponentially with the number of variables.  The number of
states ${\cal N}(E)$ with energy $E$ is written as:
\begin{equation}
{\cal N}(E) = \exp\left( N \Sigma(\alpha,\epsilon) \right)
\label{complexity_def}
\end{equation}
where the quantity $\Sigma(\alpha,\epsilon)$ is called the complexity.  It
is a function of $\alpha=M/N$ and $\epsilon=E/N$, and the form
(\ref{complexity_def}) is derived from the basic assumption that $\log
{\cal N}(E)$ is extensive.  In general, whenever a problem has a nonzero
complexity, one may expect that simple local algorithms will have great
difficulty in finding the ground state, simply because the states
proliferate (for large $N$) and the algorithm will easily get trapped
into one state with energy above that of the ground state. We shall
see in the next sections how the cavity method can handle such a
situation.

\section{A  primer on  the cavity method at zero temperature}
\label{sect_cavity}

The cavity method was originally introduced in \cite{MPV_cav} to study
spin glasses, but it gives a general framework for computing
statistical properties of various frustrated systems, and is ideally
adapted to systems with a locally tree-like structure.  It is always in
principle equivalent to the replica method, which is a more compact
and very appealing formalism, however it possesses two advantages. On
one hand, it proceeds through a standard probabilistic analysis, and
makes explicit all the hypotheses involved in it. Roughly speaking,
the cavity method assumes some properties about the correlations
between variables in a system with $N$ spins, and shows that these are
self-consistently reproduced for a system with $N+1$ spin system. The
problem in turning it into a rigorous methods is that these hypotheses
only hold in the large $N$ limit, not for $N$ small. If one is able to
have a good control of the correlations as a function of $N$, then the
cavity method becomes also a choice method for rigorous probabilistic
studies of frustrated systems \cite{talag}.  On the other hand, in the
cavity method, one considers explicitely the site dependence of the
order parameter, and the averaging over 'disorder' is performed at the
end (this is in contrast with the replica approach where the disorder
average is made from the very beginning).  As we shall see here, this
aspect allows to define some algorithm, inspired from the cavity
method, which computes the order parameter on each site for one given
sample.

In what follows we shall be interested in the zero temperature version
of the cavity method. As discussed in details in \cite{MP_T0}, the
formalism simplifies a lot in this limit.  Here we shall mainly
outline for completeness the basic aspects of the method, applied to
the 3-sat model where all function nodes involve exactly three spins
(the generalization to more general problems is totally
straightforward but would make the notation more cumbersome). We refer
the interested reader to \cite{MP_T0} for more details. We shall
first present the method in its simple 'replica symmetric' (RS)
version where it assumes the presence of a single state, and we shall
then turn to the more involved case in which many states exist but are
uncorrelated, a situation called 'one step replica symmetry breaking'
(1RSB) in the replica jargon.

\subsection{ The cavity method with one single state ('RS' case)}
\label{RScavity}
\subsubsection{Adding one spin}
\label{addonespin}

\begin{figure}
\centering
\includegraphics[width=12cm]{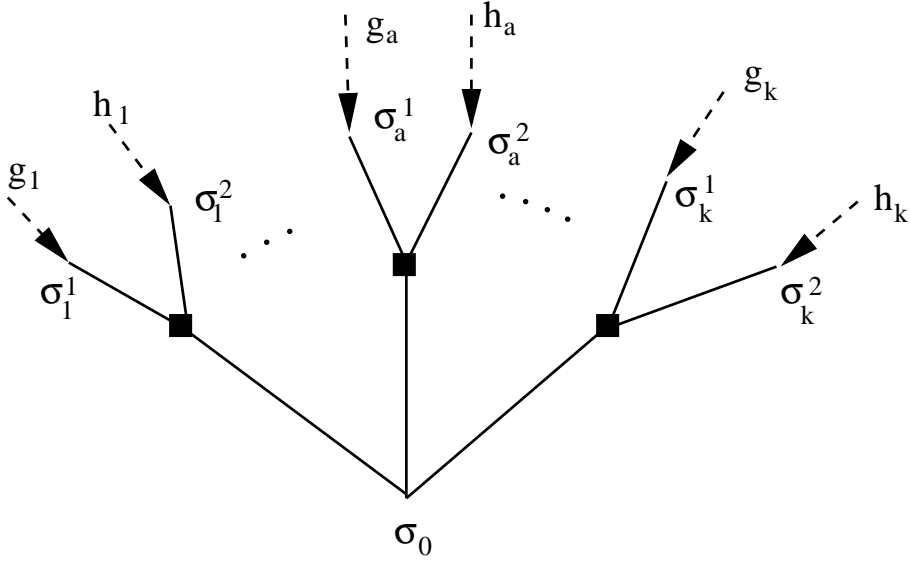}
\caption{When a new spin $\sigma_0$ is added to the system, it
gets connected through $k$ new function nodes to $2k$ other spins,
$\sigma_a^1$ and  $\sigma_a^2$. The cavity-field on $\sigma_a^1$
is denoted by $g_a$, the one on  $\sigma_a^2$ is denoted by $h_a$.
}
\label{fig_iter}
\end{figure}

Consider a $N$ spin system $\sigma_1,..,\sigma_N$ and its interaction
graph, and add to it a new spin $\sigma_0$. Then generate the new
function nodes involving this new spin as follows: for each pair $1\le
i< j\le N$, the function node $(0,i,j)$ is present with probability $6
\alpha/N^2$.  Therefore we have added $k$ new function nodes which we
label by $a=1,...,k$, where $k$ is a random variable with probability
distribution $f_{3 \alpha}(k)$.  Let us consider all the new function nodes which
involves, besides $\sigma_0$, $2k$ other spins which we call
$\sigma_a^1$ and $\sigma_a^2$ (see fig.\ref{fig_iter}). Generically,
on the original graph (i.e. before adding $\sigma_0$), these spins are
far apart from each other.  If there exists only one state, the
clustering property implies that the correlations between these spins,
before adding $\sigma_0$, vanish. Using the fact that $\sigma_{a}^1$
and $\sigma_{a}^2$ are binary variables, this decorrelation implies
that the minimal energy of the original graph, for fixed values of the
$2k$ spins $\sigma_{a}^1$ and $\sigma_{a}^2$, can be written as
\begin{equation}
	E(\{ \sigma_{a}^1,\sigma_{a}^2\} )= A
-%\frac{1}{2} 
\sum_{a=1}^k
\left( g_{a} \sigma_a^1 +h_{a} \sigma_a^2\right)
\end{equation}
where the $2k$  'local fields' $g_a$ and $h_a$ are nothing but cavity-fields passed from
each spin to the function node $a$, and $A$ is a constant (independent of the
local fields).

Looking at the function node $a$ in the full graph including spin
$\sigma_0$, we need to minimize the function $
E_a(\sigma_0,\sigma_a^1,\sigma_a^2) - ( g_{a} \sigma_a^1 +h_{a}
\sigma_a^2)$ with respect to $\sigma_a^1,\sigma_a^2$.  This is
precisely what one does in the message passing procedure of
sect.\ref{sect_sum_prod}, and one can thus use (\ref{uwdef2}) to get
the minimal energy of the new graph with $N+1$ spins, for a given
value of $\sigma_0$:
\begin{equation}
E(\sigma_0)=A-\sum_{a=1}^k  \hat w_{\bf J_a}( g_{a}, h_{a})
- \sigma_0 \sum_{a=1}^k \hat u_{\bf J_a}( g_{a}, h_{a})
\label{iterE}
\end{equation}
where $E_0=A-\sum_{a=1}^k \left(\vert h_{a}^1\vert+ \vert h_{a}^2 \vert
\right)$ is the minimal energy of the $N$ spin system.

Equation (\ref{iterE}) shows that the 'cavity field' on the new spin
$\sigma_0$ (the coefficient of $-\sigma_0$) can be written as:
\begin{equation}
h_0= \sum_{a=1}^k \hat u_{\bf J_a}( g_{a}, h_{a}) \ .
\end{equation}
As we shall see, its is often useful to decompose this cavity-field as
a sum of cavity-biases:
\begin{equation}
h_0= \sum_{a=1}^k u_a
\ \ \  ; \ \ \ 
u_a= \hat u_{\bf J_a}( g_{a}, h_{a})
\label{iterhu}
\end{equation}

\subsubsection{Self-consistency equation: the order parameter}

Whenever one adds a new spin $\sigma_0$, one picks up a value of $k$,
and a set of $2k$ fields $ g_{a}, h_{a}$, which are iid variables
taken from a probability distribution ${\cal P} (h)$.  The cavity method
assumes the existence of a thermodynamic limit $N \to \infty$ where
the energy density $E/N$ and the distribution of local fields ${\cal P}
(h)$ have well defined limits. This means that the distribution of
$h_0$ is the same as that of the $2k$ fields.  Calling ${\cal Q}(u) $ the
probability distribution of the $u$ variables, this stability
condition of the iteration(\ref{iterhu}) implies that:
\begin{eqnarray}\nonumber
{\cal Q}(u)&=&\int dg dh {\cal P}(g) {\cal P}(h) {\cal E}_J \delta\left( u-\hat u_{\bf J}(g,h)\right)
 \\
{\cal P}(h)&=&\sum_{k=0}^\infty f_{3 \alpha}(k) \int du_1...du_k {\cal Q}(u_1)...{\cal Q}(u_k) 
\delta\left(h-\sum_{a=1}^k u_a \right)
\label{iterPh}
\end{eqnarray}
where ${\cal E}_J$ means an expectation value with respect to all the couplings $J$.

\subsubsection{Computing the energy}
\label{eners}
One can easily compute the average shift in the ground state energy
when adding one new spin. Looking at the addition process defined in
sect.\ref{addonespin}, we see that the energy of the original graph
with $N$ spins is $A-\sum_{a=1}^k (\vert g_a\vert+\vert h_a \vert)$,
while that of the $N+1$ spin system is $A- \sum_{a=1}^k \hat w_{\bf
J}( g_{a}, h_{a}) - \vert \sum_{a=1}^k \hat u_{\bf J}( g_{a},
h_{a})\vert$.  Therefore the shift in energy when adding the new spin
$\sigma_0$ is
\begin{equation}
\Delta E_1 ^{(0)}= \sum_{a=1}^k \left( -\hat w_{\bf J_a}( g_{a}, h_{a})
+ \vert g_a\vert+\vert h_a \vert\right) 
- \vert \sum_{a=1}^k \hat u_{\bf J_a}( g_{a}, h_{a})\vert\ .
\label{eneshift}
\end{equation}

Equation (\ref{iterPh}) gives an integral equation for the 'order
parameter' which is the probability distribution ${\cal P}(h)$ (or
alternatively ${\cal Q}(u)$).  Let us now suppose that this equation has
been solved (we shall see below how this can be done on each specific
example), and show how the energy density can be deduced from this
order parameter.  We must compute the average of the energy shift for
adding one spin (averaged over the choice of $k$ and of the
corresponding cavity-fields):
\begin{eqnarray}
\Delta E_1&=&\sum_{k=0}^\infty f_{3 \alpha}(k) { {\cal E}_J} 
 \int \prod_{a=1}^k \left[ dg_a {\cal P}(g_a) dh_a {\cal P}(h_a)
 \right] \nonumber \\&&
 \left(\sum_{a=1}^k \left[-\hat w_{\bf J_a}( g_{a}, h_{a})
		+ \vert g_a\vert+\vert h_a \vert\right]
		- \vert \sum_{a=1}^k \hat u_{\bf J_a}( g_{a}, h_{a})\vert
\right)
\label{deltae1}
\end{eqnarray}
One might believe that, as the energy grows linearly in $N$ at large
$N$, this average energy shift would be equal to the energy density;
however there is a correction term due the change in the number of
function nodes per variable in the iteration $N \to N+1$. Indeed in
the $N+1$ spin system we are generating function nodes with
probability $6 \alpha/ N^2$ in a system with $N+1$ vertices and
therefore we are slightly over-generating function nodes. We need to
cancel a fraction $1-N^2/(N+1)^2 \simeq 2/N$ of them at random: the
probability of deleting $k'$ function nodes is
\begin{equation}
\left(
\begin{array}{c}
\alpha N \\ k'
\end{array} 
\right)
(2/N)^{k'} (1-2/N)^{\alpha N-k'}
\simeq (2 \alpha)^{k'} \exp(-2 \alpha)/{k'!} = f_{2 \alpha}(k')
\end{equation}
and the average number of deleted function nodes is $2 \alpha$.
Each deleted function node contributes to the average energy change with a
correction term:
\begin{eqnarray}
\Delta E_2&=&
 { {\cal E}_J}   \int dh_1 dh_2 dh_3 {\cal P}(h_1){\cal P}(h_2){\cal
 P}(h_3) \nonumber \\&&
\left(
\min_{\sigma_1,\sigma_2,\sigma_3}\left[
E(\sigma_1,\sigma_2,\sigma_3)
-h_1 \sigma_1-h_2 \sigma_2-h_3 \sigma_3 \right]
+\vert h_1\vert +\vert h_2\vert + \vert h_3 \vert
\right)
 \label{deltae2}
\end{eqnarray}
The ground state energy density is finally given by:
\begin{equation}
\epsilon_0=\Delta E_1 -2 \alpha \Delta E_2
\label{enetotrs}
\end{equation}

\subsection{The cavity method with many states ('one step rsb')}
\label{cavity1rsb}
\subsubsection{Iteration within one state}
Let us now see how the cavity method can be used to handle a situation
in which there exist many states. As far as the clustering condition
holds within each state, the iterative method can still be applied to
each state. The problem is that the iteration induces some crossings
of the energies of the states, and one needs to take this effect into
account properly.  We proceed as in the previous section by adding the
new spin $\sigma_0$ connected to the $2k $ spins $\{ \sigma_{a}^1,
\sigma_{a}^2\}$.  In each state $\alpha$, one can reproduce the
previous arguments: due to the vanishing of correlations, the energy
of the state $\alpha$, for fixed values of the $2k$ spins $\{
\sigma_{a}^1\}$ and $\{ \sigma_{a}^2\}$, can be written as
\begin{equation}
	E^\alpha(\{ \sigma_{a}^1,\sigma_{a}^2\} )= A^\alpha
-\sum_{a=1}^k
\left( g_{a}^\alpha \sigma_a^1 +h_{a}^\alpha \sigma_a^2\right) \ .
\end{equation}
We now have, for each state $\alpha$, $2k$ 'local fields'.

Within each state $\alpha$, the optimization procedure on the $2k$
spins $\sigma_a^1,\sigma_a^2$ proceeds as before. The minimal energy
of the new graph with $N+1$ spins, for a given value of $\sigma_0$,
is:
\begin{equation}
E^\alpha(\sigma_0)=A^\alpha-\sum_{a=1}^k  \hat w_{\bf J_a}( g_{a}^\alpha, 
h_{a}^\alpha)
- \sigma_0 \sum_{a=1}^k \hat u_{\bf J_a}( g_{a}^\alpha, h_{a}^\alpha) \ .
\label{iterE2}
\end{equation}
This  shows that the local field on the new spin $\sigma_0$ 
in state $^\alpha$
can be written as
\begin{equation}
h_0^\alpha= \sum_{a=1}^k \hat u_{\bf J_a}( g_{a}^\alpha, h_{a}^\alpha) \ ,
\label{haliter}
\end{equation}
and the shift in energy of this state is (see (\ref{eneshift})):
\begin{equation}
\Delta E^\alpha=\sum_{a=1}^k  
\left(-\hat w_{\bf J_a}( g_{a}^\alpha, h_{a}^\alpha)
+ \vert g_a^\alpha\vert+\vert h_a^\alpha \vert\right) 
- \vert \sum_{a=1}^k \hat u_{\bf J_a}( g_{a}^\alpha, h_{a}^\alpha)\vert\ .
\label{ealshift}
\end{equation}

\subsubsection{Hypotheses on the states; u-surveys}
We suppose the existence of many states, with a complexity function
$\Sigma(\alpha,E/N)$ defined as in (\ref{complexity_def}) which is an
increasing convex function. Let us consider all the states $\alpha$
with a given energy density $E/N=e$. We suppose that all the local
fields $h_{j \to a}^\alpha$ on a given edge $j \to a$ are iid, taken
from a probability distribution $P_{j \to a} ^{e} (h)$ called a
h-survey.  This probability distribution fluctuates from one edge to
the next, so that the full order parameter, obtained by averaging over
edges, is the functional probability distributions of these h-surveys.
The same hypotheses hold for the distribution of the cavity-biases:
all the $u_{a \to 0}$ on a given link are iid taken from a probability
distribution $Q_{a \to 0} ^{e} (u)$ called a u-survey.  Notice that
the previous 'RS' solution corresponds to having deterministic
messages on each edge $P_{j \to a} ^{e} (h) =\delta(h-h_{j \to a})$
and $Q_{a \to 0} ^{e} (u)= \delta (u- u_{a \to 0})$. In the many state
hypothesis, we have to generalize the messages, and the important
quantities are the probability distributions (over the many states
having a fixed energy density) of the cavity-biases going through a
given link.

\subsubsection{Iteration: level crossings and reweighting}
One iteration step of the cavity procedure leads to an equation
relating the probability distributions, before any averaging over the
graph. In our iteration procedure, the h-survey on the new site,
$P_0^e(h)$, is related to the h-surveys on the other $2 k$ spins
$\sigma_a^1,\sigma_a^2$ computed in the absence of $\sigma_0$.  Let us
denote by $P_a^e(g_a) $ the h-survey incoming onto $\sigma_a^1$ and
${P_a^e} '(h_a) $ the h-survey incoming onto $\sigma_a^2$. The
h-survey $P_0^e(h)$ is given by: 
\begin{eqnarray} P_0^e(h)&=& C\int
\prod_{a=1}^k\left[P_a^e(g_a) dg_a \; {P_a^e} '(h_a) dh_a\right]
\delta\left[h-\sum_{a=1}^k \hat u_{\bf J_a}(g_a,h_a)\right] \nonumber \\ 
&& \exp\left[y\sum_{a=1}^k \left(\hat w_{\bf J_a}(g_a,h_a)-\vert g_a \vert
-\vert h_a \vert\right)+y \vert \sum_{a=1}^k \hat u_{\bf J_a}(g_a,h_a)
\vert\right]
\label{iterbasic}
\end{eqnarray}
where $C$ is a normalisation constant insuring that $P_0^e(h)$ is a normalised
probability distribution.
This equation provides the generalization to the RSB case of the
simple iteration (\ref{iterhu}) of the previous section. Two
complications have appeared: the simple messages (cavity-fields and
cavity-biases) have become h-surveys, i.e. probability distributions
of simple messages, and a new term has appeared which is the
exponential reweighting term. In this term, the parameter $y$ is a
number equal to the derivative of the complexity with respect to the
energy.
\begin{equation}
y=\frac{ \partial\Sigma}{\partial e}
\end{equation}
Let us now explain the origin of this new and crucial term.  For a
given state $\alpha$, we add one new spin to the system and want to
compute the new h-survey.  In this process, we have seen in
(\ref{ealshift}) that there is an energy shift $\Delta E^\alpha$ {\it
which depends on the state, and is correlated to the value of the
cavity-field $h=\sum_{a=1}^k \hat u_{\bf J}( g_{a}^\alpha,
h_{a}^\alpha) $}.  Let us call $S_0(h,\Delta E)$ the joint probability
(when looking at all states) of the cavity-bias and the free energy
shift, for this function node.

When we compute $P_0^e(u)$ at a fixed energy density $e=E/N$, we get a
contribution from all states with energies before iteration equal to
$E-\Delta E$.  Therefore:
\begin{equation}
P_0^e(h)=C \int d(\delta E) S_0(h,\Delta E)
\exp\left(N\Sigma\left(\frac{E-\Delta E}{N}\right)\right)\simeq C' \int d(\delta E)
S_0(h,\Delta E) \exp(-y \Delta E) \ .
\end{equation}
The reweighting term in $\exp(-y \Delta E)$ is due to the level
crossing, and the fact that the complexity $\Sigma(\alpha,\epsilon)$ is
not constant, but increasing: states with a negative value of the
energy shift are thus favoured.

It is useful, as before, to decompose the iteration procedure
(\ref{iterbasic}) into two steps and introduce the u-surveys. On any
function node $a$, we merge two h-surveys $P_a^e(g_a)$ and ${P_a^e}
'(h_a)$ in order to build a u-survey:
\begin{equation}
Q_{a }^e(u)=\int dg dh P_{a}^e(g) {P_{a}^e} '(h) 
 \delta\left( u-\hat u_{\bf J_a}(g,h)\right) 
 \exp \left( y \left[\hat w_{\bf J_a}(g,h)-\vert g \vert -\vert h \vert \right]\right)
\label{iterQu_rsb}
\end{equation}
Then we can combine all the u-surveys incoming onto the new spin in
order to build its h-survey:
\begin{equation}
P_{0}^e(h)=\int du_1...du_k \;
Q_{1}^e (u_1)...Q_{k}^e(u_{k}) \; 
\exp\left(y\vert \sum_{a=1}^k u_a\vert\right) \;
\delta\left(h-\sum_{a=1}^{k} u_a \right) \ .
\label{iterPh_rsb}
\end{equation}
Note that there is some degree of arbitrariness in the way one
distributes the reweighting between the two iteration steps: different
choices amount to different definitions of u-surveys. The above one is
the most natural one, and this is what we shall adopt from now on.

\subsubsection{Order parameter and self-consistency: population dynamics}
\label{sect_popdyn}
Equations (\ref{iterQu_rsb},\ref{iterPh_rsb}) are the main result
giving the way to compute the messages sent along to a new added
site. The assumed existence of a thermodynamic limit allows in
principle to write a self consistency of the iteration in a way
similar to (\ref{iterPh}). In the present case, this is an equation
for the functional ${\cal P}(P(h))$ giving the probability, when one picks
up an edge $i \to a$ at random, to observe on this edge a h-survey
$P_{i \to a} (h)$ equal to $P(h)$.  Alternatively, one can use the
functional ${\cal Q}(Q(u))$ giving the probability, when one picks up an
edge $a \to j$ at random, to observe on this edge a u-survey $Q_{a \to
j} (u)$ equal to $Q(u)$.  In the following we shall rather work with
the u-surveys which turn out to have a simpler structure in practice,
but obviously a fully equivalent description can be obtained working
with h-surveys.

These functional equations are the generalisation to the RSB case of
the equations (\ref{iterPh}) for the RS case. One could write them
explicitely, but they are not particularly illuminating, and we prefer
to work directly with the iteration equations
(\ref{iterQu_rsb},\ref{iterPh_rsb}). These define a stochastic
process; at each iteration, one performs the following operations:
\begin{itemize}
\item
Pick up at random a number of neighbours $k$, with the probability
$f_{3 \alpha}(k)$;
\item
Pick up at random $k$ u-surveys $Q_1(u_1),...,Q_k(u_k)$ from the
distribution ${\cal Q}(Q(u))$;
\item
Compute a h-survey $P_1(g)$ as the reweighted convolution 
\begin{equation}
P_1(g)= C_1\int du_1...du_k
Q_1(u_1)...Q_k(u_k)
\exp\left(y \vert \sum_{a=1}^k u_a \vert \right) \;
\delta\left(g-\sum_{a=1}^k u_a \right) \ .
\label{p1def}
\end{equation}
\item
Pick up at random a number of neighbours $k'$, with the probability
$f_{3 \alpha}(k')$;
\item
Pick up at random $k'$ u-surveys $Q_{k+1}(u_1),...,Q_{k+k'}(u_{k'})$
from the distribution ${\cal Q}(Q(u))$;
\item
Compute a h-survey $P_2(h)$ as the convolution 
\begin{equation}
P_2(h)= C_2 \int du_1...du_k
Q_{k+1}(u_1)...Q_{k+k'}(u_{k'})
\exp\left(y \vert \sum_{a=1}^{k'} u_a \vert \right) \;\delta\left(h-\sum_{a=1}^{k'} u_a \right) \ .
\label{p2def}
\end{equation}
\item
Pick up at random a set of couplings ${\bf J}$ caracterising a new function
node, from the a priori distribution of couplings.
\item
Compute a new u-survey, $Q_0(u)$ as 
\begin{equation}
 Q_0(u)= C_0 \int  dg dh P_1(g) P_2(h) 
 \delta\left( u-\hat u_{\bf J}(g,h)\right) \exp \left( y \left[\hat w_{\bf J}(g,h)
-\vert g \vert -\vert h \vert \right] \right)\ , 
\label{Qiterpop}
\end{equation}
\end{itemize}
where $C_0$ is a normalisation constant insuring that $Q(u)$ has an
integral equal to one.

This iteration defines a stochastic process in the space of u-surveys,
which in turn defines a flow for ${\cal Q}(Q(u))$, of which we would like
to compute the fix point. Following \cite{MP_Bethe}, this in done in
practice by a population dynamics algorithm: one uses a representative
population of ${\cal N}$ u-surveys from which the various
$Q_\ell(u),\ell\in\{1,...,k+k'\} $ used in the iteration are
extracted.  After $Q_0(u)$ has been computed, one of the u-surveys in
the population, chosen randomly, is erased and substituted by
$Q_0(u)$. After some transient, this population dynamics algorithm
generates sets of u-surveys which are sampled with a frequency
proportional to the seeked ${\cal Q}(Q(u))$.

The point of this stochastic process approach is to avoid trying to
write explicitely the complicated functional equation satisfied by
${\cal Q}(Q(u))$. This is one crucial place where the cavity method turns
out to be superior to the replica method: with replicas one performs
the average over disorder from the beginning, and one is forced to
work directly with the functional ${\cal Q}(Q(u))$ \cite{Monasson_func}.  As
this is very difficult, people have thus been constrained to look for
approximate solutions of ${\cal Q}(Q(u))$ where the functional is taken in
a simple subspace, allowing for some explicit computations to be done.

\subsubsection{Computing the energy and the complexity}
Here we show how to generalize the computation of the energy of
sect.\ref{eners} to the 1RSB case.

When adding one site $0$, connected through $k$ function nodes to $2k$
 sites the energy shift in one given state is:
\begin{equation}
\delta E=  \sum_{\ell=1}^k\left( -\hat w_{\bf J_\ell}(h_\ell,g_\ell)+\vert h_\ell\vert
+\vert g_\ell \vert\right)
-\vert \sum_{\ell=1}^k
\hat u_{\bf J_\ell}(h_\ell,g_\ell) \vert \ ,
\label{de_def}
\end{equation}
where $h_\ell,g_\ell$ are the incoming fields onto the function node
number $\ell$. Let us call $P_\ell(h_\ell)$ and $P_\ell'(g_\ell)$ the
corresponding field distributions (the h-surveys). They induce a
probability distribution $P_0(\delta E) $ of the energy change
(\ref{de_def}):
\begin{equation}
P_0(\delta E)= \int \prod_{\ell=1}^k \left[dh_\ell dg_\ell \;
P_\ell(h_\ell) P_\ell'(g_\ell)\right]\delta
\left(
\delta E
+ \sum_{\ell=1}^k \left[\hat w_{\bf J_\ell}(h_\ell,g_\ell)-\vert h_\ell\vert
-\vert g_\ell\right]  
+\vert \sum_{\ell=1}^k
\hat u_{\bf J_\ell}(h_\ell,g_\ell) \vert 
\right)
\ .
\label{P0_def}
\end{equation}
Let us look at the corresponding change in complexity.  The new system
has $N+1$ variables and $M+k$ function nodes, and its number of states
at energy $E$, $\exp((N+1)\Sigma([M+k]/[N+1],[E/N+1]))$ is given by:
\begin{equation}
\exp\left[(N+1)  \Sigma \left(\frac{M+k}{N+1},\frac{E}{N+1}\right)\right]
=\int d(\delta E) \;
P_0(\delta E) \exp\left[N \Sigma \left(\frac{M}{N},\frac{E-\delta E}{N}\right) \right]
\ .
\label{above}
\end{equation}
This expression depends on the precise spin which has been added
through the choice of the distributions $P_\ell$ and $P_\ell'$, and of
couplings ${\bf J_\ell}$, which appear in (\ref{de_def}). As one
expects $\Sigma(\alpha,\epsilon)$ to be self-averaging, one must average
the logarithm of the expressions in (\ref{above}) over the iteration
of population dynamics algorithm.  We denote this averaging by an
overline.  As $\delta E$ is finite, one can expand in $\delta E /N$ in
the thermodynamic limit, to get (calling as always $\alpha=M/N$ and
$\epsilon=E/N$):
\begin{equation}
\Sigma(\alpha,\epsilon)-\epsilon \frac{\partial\Sigma}{\partial \epsilon}+2 \alpha
\frac{\partial\Sigma}{\partial \alpha}
=\overline{\log\left(\int d(\delta E) \;
P_0(\delta E) \exp\left[-y \delta E\right]\right)} \ ,
\label{shift_comp1}
\end{equation}
where we used the fact that $\overline{k}=3 \alpha$. As in the RS
case of (\ref{deltae2}), the derivative ${\partial\Sigma}/{\partial
\alpha}$ can be computed by adding one function node to the system.
For a generic function node $a$ connected to the sites $1,2,3$ and
with interaction coupling ${\bf J}$, the probability distribution
$P_a(\delta E) $ of the energy change is:
\begin{eqnarray}
P_a(\delta E)
&=& \int dh_1 dh_2 dh_3 P_1(h_1) P_2(h_2) P_3(h_3) \nonumber \\&&
\delta\left[ \delta E -\min_{\sigma_1,\sigma_2,\sigma_3}\{-h_1 \sigma_1 
-h_2 \sigma_2- h_3 \sigma_3 +\epsilon_{\bf J} (\sigma_1, \sigma_2, \sigma_3)\}
-\left(\vert h_1\vert +\vert h_2\vert + \vert h_3 \vert\right)\right] \;
\label{Pa_def}
\end{eqnarray}
Let us look at the corresponding change in complexity.
The new system has $N$ variables and $M+1$ function nodes, so that:
\begin{equation}
\exp\left[N  \Sigma \left(\frac{M+1}{N},\frac{E}{N}\right)\right]
=\int d(\delta E) \;
P_a(\delta E) \exp\left[N \Sigma \left(\frac{M}{N},\frac{E-\delta E}{N}\right)\right] \ .
\label{above2}
\end{equation}
After averaging over the iterations, this gives:
\begin{equation}
\frac{\partial\Sigma}{\partial \alpha}
=\overline{\log\left(\int d(\delta E) \;
P_a(\delta E) \exp\left[-y \delta E\right]\right)} \ .
\label{shift_comp2}
\end{equation}
Combining the two expressions (\ref{shift_comp1},\ref{shift_comp2}),
it turns out that the quantity which is computed naturally in this
scheme is the Legendre transform $\Phi(y)$, with respect to the energy
density $\epsilon$, of the complexity 
function $\Sigma(\alpha,\epsilon)$ \cite{Mon95,FraPar}.
This 'zero temperature free energy' is defined precisely as:
\begin{equation}
\Sigma(\alpha,\epsilon)-y \epsilon=- y \Phi(y) \ \ \ ; \ \ \ y \equiv \frac{d \Sigma}{d \epsilon}
\label{legendre}
\end{equation}
and it can be computed from the population dynamics as:
\begin{eqnarray}
\Phi(y)&=& \Phi_1(y)-2 \alpha \Phi_2(y)\\
 \Phi_1(y)&=&-\frac{1}{y}\overline{\log\left(\int d(\delta E) \;
P_0(\delta E) \exp\left[-y \delta E\right]\right)}\nonumber \\ 
\Phi_2(y)&=&-\frac{1}{y}\overline{\log\left(\int d(\delta E) \;
P_a(\delta E) \exp\left[-y \delta E\right]\right)} \ ,
\label{PHIrsb}
\end{eqnarray}
where $P_0$ and $P_a$ are given in (\ref{P0_def},\ref{Pa_def}).

Technically, it turns out that $\Phi_1(y)$ is more easily computed
through the normalisation of the u-surveys. When we compute a u-survey
$Q_0(u)$ as in (\ref{Qiterpop}), we can memorize the corresponding
normalisation constant $C_0$. Picking up $k$ u-surveys $Q_\ell$ at
random in the population, and calling $C_\ell$ the corresponding
normalizations, one gets:
\begin{equation}
\Phi_1(y) = -\frac{1}{y}\overline{\log\left(\int \prod_{\ell=1}^k \left[du_\ell 
\frac{Q_{\ell }(u_\ell)}{C_{\ell } } \right] \; 
 \exp\left[{y \vert \sum_{\ell=1}^k u_\ell \vert}\right]\right)}
\end{equation}

\section{The cavity method applied  to the random 3-sat problem}
\label{sect_r3sat}

\subsection{Known results on the phase diagram}
Considering the random 3sat problem where the graph is generated at
random and the various couplings take values $\pm 1$ with probability
$1/2$, numerical experiments have provided a detailed study of the
probability $P_N (\alpha , K)$ that a given ${ F}$ including $M=\alpha
N$ clauses be satisfiable. For large sizes, there appears a remarkable
behaviour~: $P$ seems to reach unity for $\alpha < \alpha _c (K)$ and
vanishes for $\alpha > \alpha _c (K)$ \cite{AI_issue}.  Such an abrupt
threshold behaviour, separating a SAT phase from an UNSAT one, has
indeed been rigorously confirmed for 2-SAT, which is in P, with
$\alpha _c (2) =1$ \cite{2SAT_1,2SAT_2,2SAT_3}.  For larger $K \ge 3$, K-SAT
is NP-complete and much less is known.  The existence of a sharp
transition has not been rigorously proven yet but estimates of the
thresholds have been found. The present best numerical estimate for
$\alpha_c$ at $K=3$ is $4.26$~\cite{Crawford}, and the rigorous bounds
are\cite{bounds_1,bounds_2,bounds_3,bounds_4} $3.26<\alpha_c<4.506$ , while previous
statistical mechanics analysis using the replica method, has found
$\alpha_c(3)\sim 4.48$~\cite{BMW} and $\alpha_c(3)\sim
4.396$~\cite{FLRZ} in the framework of variational approximations.

The interest in random K-SAT arises from the fact that it has been
observed numerically that hard random instances are generated when the
problems are critically constrained, i.e. close to the SAT/UNSAT phase
boundary \cite{AI_issue,MZKST}. The study of such hard instances
represent a theoretical challenge towards a concrete understanding of
complexity and the analysis of algorithms\cite{TCS_issue}. Moreover,
hard random instances are also a test-bed for the optimization of
heuristic (incomplete) search procedures, which are widely used in
practice.

\subsection{The cavity analysis with one state}

In the 3sat problem, the energy of a function node, as given by
(\ref{epsdefKsat}), is
\[
E_a= \left(1+J_1^a \sigma_{1}\right)
\left(1+J_2^a \sigma_{2}\right)\left(1+J_3^a \sigma_{3}\right)/4
\]
 and leads to:
\begin{eqnarray}
\hat w_{\bf J_a}(h_2,h_3)&=& |h_2|+|h_3| 
-\theta(J_2^a h_2) \theta(J_3^a h_3) 
\nonumber \\ \hat u_{\bf J_a}(h_2,h_3)&=& -J_1^a \theta(J_2^a h_2)
\theta(J_3^a h_3 ) \ ,
\label{wu_Ksat}
\end{eqnarray}
where the function $\theta(x)$ is defined as 
\begin{equation}
\theta(x)= 1 \ \ \mbox{if} \ \ x>0\ \ \ \ ; \ \ \ \ \theta(x)= 0 \ \ 
\mbox{if} \ \ x\le 0
\end{equation}

Let us consider the cavity iteration scheme, within the hypothesis of
there being a single state. We thus use the general formalism
presented in sect.  \ref{RScavity}. In the case of 3sat,
(\ref{wu_Ksat}) shows that the cavity-bias on a given edge $u_{a \to
j}$ takes values either in $0,1$ if the corresponding coupling is
negative, otherwise it takes values in $-1,0$.  Therefore the
cavity-fields are integers. (This is the reason for using the unusual
factor $2$ for each violated clause in the definition
(\ref{epsdefKsat}) of the energy). Because the function $\hat u_{\bf
J_a}$ is an odd function of one of the couplings in ${\bf J_a}$, and
these couplings are random variables taking values $\pm 1$ with
probability $1/2$, the distribution ${\cal Q}(u)$ of cavity-biases must be
of the form:
\begin{equation}
{\cal Q}(u)= c_0 \delta(u)+\frac{(1-c_0)}{2} (\delta(u-1)+\delta(u+1)) .
\label{Qc0_3sat}
\end{equation}
Plugging this expression into the self-consistency equations
(\ref{iterPh}) leads to a relation between $c_0$ and the weight
$p_0\equiv{\cal P}(h=0)$:
\begin{equation}
p_0= \sum_k f_{3 \alpha}(k) \sum_{q=0}^{[k/2]}
\left( 
\begin{array}{c}
k \\ 2 q
\end{array} 
\right) 
c_0^{k-2 q} (\frac{1-c_0}{2})^{2 q} 
\left( 
\begin{array}{c}
2 q \\ q
\end{array} 
\right) 
=\exp(-3\alpha(1-c_0))
I_0(3\alpha(1-c_0)) \ ,
\label{p0_3sat}
\end{equation}
where $I_0$ is the Bessel function and $[k/2]$ is the integer part of $k/2$.
Let us now compute $c_0$. From (\ref{wu_Ksat}) we find that a cavity-bias
vanishes whenever at least one of the incoming fields ($h_2$ or $h_3$) is zero
or has a sign opposite to the corresponding coupling. This shows that
\begin{equation}
c_0=1 - \operatorname{Prob}[(h < 0)\cap (g < 0)]=1-(\frac{1-p_0}{2})^2
\label{C0_3sat}
\end{equation}
We obtain a closed set of equations (\ref{C0_3sat},\ref{p0_3sat})
which is easily solved. The distribution ${\cal P}(h)$ of
cavity fields is then given by ${\cal P}(h)=\sum_r p_r\delta(h-r)$,
where the weights $p_r$ are equal to:
\begin{equation}
p_r\equiv {\cal P}(h=r)={\cal P}(h=-r)=\sum_{k=r}^\infty f_{3 \alpha}(k)
\sum_{q=0}^{[\frac{k-r}{2}]} 
\left( 
\begin{array}{c}
k \\ 2 q+r
\end{array} 
\right) 
c_0^{k-2 q-r}
(\frac{1-c_0}{2})^{2 q+r} 
\left( 
\begin{array}{c}
2 q + r \\ q
\end{array} 
\right) 
\ ,
\label{pr}
\end{equation}
 and the energy
is computed from  (\ref{deltae1},\ref{deltae2}).

A non-trivial solution exists for $\alpha> 4.667$, with a ground state
energy that becomes positive at $\alpha=5.18$.  The prediction of this
hypothesis assuming a single state is a paramagnetic SAT phase with
$c_0=p_0=1$, and energy $E_0=0$ for $\alpha< 5.18$, and a frozen UNSAT
glassy phase with $c_0<1$ and $E_0>0$ for $\alpha> 5.18$.  It is known
\cite{MoZe_prl,MoZe_pre} that this solution is wrong both
quantitatively (the location of the transition point) and
qualitatively (the structure of the order parameter).

The true transition is much more sophisticated, and the many state
formalism corresponding to 1RSB is needed to unveil its structure.

\subsection{The cavity analysis with many states}

We introduce as before the h-surveys and u-surveys,
and we use the population dynamics algorithm defined in
sect.\ref{sect_popdyn}.  It turns out to be more convenient to work
only in terms of u-surveys. The algorithm computes at each iteration a
new u-survey $Q_0(u)$ by taking $k+k'$ u-surveys
$Q_1(u),...,Q_{k+k'}(u)$ in the population through:
\begin{eqnarray}
Q_0(u) &=&
 C_0 \int \int  du_1 Q_{1}(u_1)...du_k Q_{k}(u_k)
d{v}_1 Q_{k+1}({v}_1)...d{v}_{k'} Q_{k+k'}({v}_{k'}) \nonumber \\
&&\delta\left(u-\hat u_{\bf J}(u_1+...+u_k,v_1+...+v_{k'})\right)
\exp[y \hat w_{\bf  J}(u_1+...+u_k,v_1+...+v_{k'}) ] \ .
\label{QRSBequation_3sat}
\end{eqnarray}
Here, $k$ and $k'$ are two iid random numbers taken from the Poisson
distribution $f_{3\alpha}(k)$ defined in (\ref{Poisson}), and ${\bf
J}$ denotes a set of three iid random numbers $J_1,J_2,J_3$, each
being equal to $\pm 1$ with probability $1/2$.

The functions $ \hat u$ and $\hat w$ are defined in (\ref{wu_Ksat}).
A u-survey always takes the simple form $Q_0(u)=(1-c) \delta(u) +c
\delta(u+J_1)$, it is thus a probability distribution which can be
characterized by a single number $c$, and therefore the iteration of
the population dynamics is easily done numerically.

\subsection{Solution of the self-consistency equations}
Apart from the RS solution with $y=0$ and $Q_j(u)=\delta(u-u_j)$,
 where the $u_j$ are iid taken from a distribution ${\cal Q}(u)$, the
 numerical solution finds one other solution in the region $\alpha
 >\sim 4.$ Generically, the u-surveys found can be of three types:
\begin{equation}
Q_i(u)= \left\{
\begin{array}{cc}
 \delta(u)  \; \; \; \; & \; \; \; \text{('trivial' or type a)}   \\
 (1-\eta_i e^{-y}) \delta(u)+\eta_ie^{-y}\delta(u-1) \; \; \; \; & \; \;
  \; \text{ (type $b_+$)}   \\
(1-\eta_i e^{-y}) \delta(u)+\eta_i e^{-y}\delta(u+1) \; \; \; \; & \; \;
  \; \text{ (type $b_-$)}
\end{array}
\right.
\label{Ansatz_3sat_2}
\end{equation}
The arbitrary factor $e^{-y}$ has been introduced for convenience
because numerical simulations show that the weight in $u=\pm 1$ of the
non trivial u-surveys scale proportionally to $e^{-y}$ at large $y$.
%M So that eta is like the \hat w of the notes

The statistical symmetry of the problem due to the fact that the
couplings take values $\pm 1$ with probability $1/2$ implies that the
probability of finding in the population a ``$b_+$'' message is equal
to that of finding a ``$b_-$'' message. We call $(1-t)/2$ these
probabilities, and $t$ the probability of finding a type $a$
message. Non trivial messages are fully characterized by the
distribution $\rho(\eta)$ of the $\eta_i$ variables. For any $y$, the
full solution of the problem is given by the value of $t$ and of the
function $\rho(\eta)$. It can be obtained numerically by averaging
over many iteration of the population dynamics. We shall now show how
one can get some analytic control in the large $y$ limit.

\subsubsection{Existence of non trivial u-surveys}
Looking at the iteration equation (\ref{QRSBequation_3sat}), the only
way one can obtain a trivial u-survey $Q_0(u)=\delta(u)$ is when
either $J_2 \sum_{i=1}^k u_i \leq 0 $ in the whole integration domain,
or $J_3\sum_{j=1}^{k'} v_j \leq 0$ in the whole integration domain, or
both. The probability to have $J_2 \sum_{i=1}^k u_i \leq 0 $ in the
whole integration domain is:
\begin{eqnarray}
\operatorname{Prob}\left[J_2\sum_{i=1}^k u_i \leq 0\right]&=&
\sum_{k=0}^\infty f_{3 \alpha}(k) \left( t^k+\frac{1}{2}
\frac{(1-t)}{2} t^{k-1} 2+ 2 \frac{1}{2} \left(\frac{1-t}{2}\right)^2
t^{k-2} 
\left( 
\begin{array}{c}
k \\ 2
\end{array} 
\right) 
+...
\right) \nonumber \\
&=& \exp\left[-3 \alpha \frac{1-t}{2}\right] \ .
\end{eqnarray}
The first term $t^k$ corresponds to all $Q_i(u)$, $i \in \{1,...,k\}$,
being of type $a$.  The second term corresponds to: (one of the
$Q_i(u)$ of type $b_+$, all the other ones of type $a$, and $J_2=-1$)
or (one of the $Q_i(u)$ of type $b_-$, all the other ones of type $a$,
and $J_2=+1$). The rest of the series is easily obtained similarly.

The probability $t$ of having a trivial u-survey is thus:

\begin{equation}
t=1-\left(1-\operatorname{Prob}\left[J_2\sum_{i=1}^k u_i \leq
0\right]\right)
\left(1-\operatorname{Prob}\left[J_3\sum_{j=1}^{k'} v_j \leq
0\right]\right)
=1-\left(1-\exp\left[-3 \alpha \frac{1-t}{2}\right]\right)^2
\label{EQ_t}
\end{equation}
For $\alpha$ small the only solution is the paramagnetic one $t=1$.  A
solution different from $t=1$ appears above $\alpha_0=1.63694$ (which
corresponds to $t=0.04883$).  (In fact there appears a pair of
nontrivial solutions, the relevant one is the one with lowest $t$, as
can be seen from its behaviour at large $\alpha$, and also checked in
the population dynamics algorithm).  It is interesting to observe that
this threshold coincides with the point where the so called ``unit
clause literal'' \cite{Frieze} algorithm ceases to converge.  At large
values of $\alpha$, the fraction of type $a$ u-surveys becomes very small,
e.g. $t=0.0051,0.0011,0.00025,$ at $\alpha=4,5,6$ respectively.

\subsubsection{Expansion at large $y$: location of the phase transition}
Here we shall show that, at large $y$, the scaling in $e^{-y}$ of the
weights at $u=\pm 1$ for nontrivial u-surveys is indeed consistent
with the iteration of the population dynamics
(\ref{QRSBequation_3sat}), and we deduce from this iteration a
self-consistent equation for $\rho(\eta)$.

Let us study one given iteration of (\ref{QRSBequation_3sat}).  The
probability of having $k$ cavity-biases $Q_1,...,Q_k$, among which $m$
are of type $b_+$, $n$ of type $b_-$ and $k-m-n$ of type $a$ is:
\begin{equation}
C_{k,m,n}= f_{3\alpha}(k) \frac{k! }{m!
n! (k-m-n)!}\; t^{k-m-n} (\frac{1-t}{2})^{m+n}
\label{Ckmn_def}
\end{equation}
Without loss of generality, we can assume that the $m+n$ non-trivial
u-surveys are $Q_1,...,Q_{m+n}$, and the couplings are
$J_1=J_2=J_3=1$ for this iteration. We denote by ${\bf \eta}$ the
vector of weights ${ \bf \eta} = (\eta_1,...,\eta_{m+n})$.  Since
$\hat u_{\bf J}(u_1+...+u_k,v_1+...+v_{k'})=1$ if and only if $ (J_1
\sum_i u_i>0) $ and $ (J_2 \sum_j v_j >0)$, the new u-survey $Q_0(u)$
depends only on the probabilities
\begin{equation}
f_q^{(m,n)}({\bf \eta} )\equiv \operatorname{Prob}[u_1+...+u_k=q]
\end{equation}
and is given by:
\begin{eqnarray}
Q_0(u)&=& C_0
\left\{  
  \delta(u-1) 
  \left[ \sum_{q=1}^k \sum_{q'=1}^{k'}
    f_q^{(m,n)}({\bf \eta}) f_{q'}^{(m',n')}({\bf \eta'})
    e^{y (|q|+|q'|-1)} 
  \right] 
\right.
\nonumber \\ 
&+& \left.
\delta(u) 
\left[ 
\left( \sum_{q=1}^{k} \sum_{q'=-k}^0 
+ \sum_{q=-k}^0\sum_{q'=1}^{k'}  +\sum_{q=-k}^0 \sum_{q'=-k}^0
\right)  
f_q^{(m,n)}({\bf \eta}) f_{q'}^{(m',n')}({\bf \eta'}) e^{y(|q|+|q'|)}
\right] 
\right \}
\nonumber \\
&\equiv&C_0
\left\{ 
\delta(u-1) A_0 e^{-y}+\delta(u) B_0
\right\} \ ,
\label{ABdef}
\end{eqnarray}

Iterating the population dynamics, one finds that $\rho(\eta)$
satisfies the equation
\begin{eqnarray}
\rho(\eta_0) &=& \frac{1}{1-t} \sum_{k=0}^\infty \sum_{m=1}^{k-1} \sum_{n=0}^{k-m}
C_{k,m,n}  \sum_{k'=0}^\infty \sum_{m'=1}^{k'-1} \sum_{n'=0}^{k'-m'}
C_{k',m',n'}  \times \nonumber \\
&\times & \int \prod_{\ell=1}^{m+n}\left[d\eta_\ell\rho(\eta_\ell)\right]
 \prod_{\ell'=1}^{m'+n'}\left[d\eta_{\ell'}\rho(\eta_{\ell'})\right]
\delta\left(\eta_0-\frac{A_0}{A_0 e^{-y}+B_0}\right) 
\label{EQ_RHO1}
\end{eqnarray}
This is an exact self consistent equation for the distribution
$\rho(\eta)$. It can be simplified at large $y$ , and we show in the
appendix A how to write a more tractable self-consistent equation in
this limit. This equation is best written in terms of the probability
distribution function $S(\phi)$ of the variable
$\phi=\log(1+\eta)$. One finds that $S(\phi)$ satisfies:
\begin{eqnarray}
S(\phi)&=&\int dx_+ dx_- dx_+'dx_-' A(x_+)B(x_-) A(x_+')B(x_-') \\
&&
\delta\left(\phi-\log\left[1+ \frac{(e^{x_+}-1)(e^{x_-}-1)}
{(e^{x_+}-1) e^{x_-'}+(e^{x_+'}-1) e^{x_-}+e^{x_-}e^{x_-'}}\right]\right) \ ,
\label{sphi1}
\end{eqnarray}
where $A(x)$ and $B(x)$ are two probability distributions related to
$S(\phi)$ through its Fourier transform $\hat S(q)\equiv \int d\phi
\exp(iq\phi) S(\phi)$:
\begin{eqnarray}
A(x) &\equiv& \frac{1}{e^{3 \alpha (1-t)/2}-1} \int \frac{dq}{2
\pi}  e^{-i q x} 
\left( \exp
\left[\frac{3\alpha}{2}(1-t) \hat S(q)\right]
-1 \right) \nonumber \\
B(x)&\equiv& \frac{1}{e^{3 \alpha (1-t)/2}}  \int \frac{dq}{2
\pi}  e^{-i q x}  \exp\left[3\alpha \left(t-1+\frac{1}{2}(1-t)\hat S(q)\right)\right]
\label{sphi2}
\end{eqnarray}
Equations (\ref{sphi1},\ref{sphi2}) are simple enough to  
be solved numerically to high accuracy.

Once $S(\phi)$ (and therefore $\rho(\eta)$) is known, one
can deduce the value of the zero temperature free 
energy $\Phi(y)$. To leading order
at large $y$, we show in the appendix A that $\Phi(y)=\Psi/y$,
with:
\begin{eqnarray}
\Psi&=&-\int dx dz B(x) B(z) \log\left(e^x+e^z-1\right) \nonumber \\
&&-3 \alpha \int \prod_{i=1}^2 dx_i dy_i B(x_i) B(z_i) 
\log \left[ \prod_{i=1}^2 (e^{x_i}+e^{z_i}-1)-\prod_{i=1}^2 (e^{x_i}-1) \right]
\nonumber \\
&&+2 \alpha 
  \int \prod_{i=1}^3 dx_i dy_i B(x_i) B(z_i) 
\log \left[ \prod_{i=1}^3 (e^{x_i}+e^{z_i}-1)-\prod_{i=1}^3 (e^{x_i}-1) \right]
\label{Phifinal}
\end{eqnarray}

In order to compute the $S(\phi),A(x),B(x)$ functions, solutions of
eqs (\ref{sphi1},\ref{sphi2}), we have used the fact that these
functions are probability distributions, and we have developed a
population dynamics algorithm which follows a population of $N$
variables $\phi_1,...,\phi_N$, for a given value of $\alpha$.

\begin{itemize}
\item 1) Compute $t$, the solution of $t=1-(1-\exp[3\alpha(t-1)/2])^2$.

\item 2) Initialize the $\phi_j$ as iid random positive variables, for instance with an
exponential distribution of width $1$. Initialize the 'time variable'
$\tau=1$.

\item 3) Upgrade the time  $\tau \to \tau+1$.

\item 4) Generate an integer $k \ge 1$ with the distribution 
$ \gamma^k/k! \ 1/(\exp(\gamma) -1)$, where $\gamma=3 \alpha (1-t)/2$.
 Pick up $k$ integers $i_1,..,i_k$ at random in $\{1,...,N\}$,
and compute the sum $\phi_{i_1}+...+\phi_{i_k}$. The distribution of the
variable $x_+(\tau)=\phi_{i_1}+...+\phi_{i_k}$  is $A(x_+)$, 
related to $S(\phi)$ through (\ref{sphi2}).

\item 5) Generate a random variable $x_-(\tau)$, which will be distributed according to
$B(x_-)$, as follows: with probability $\exp(-\gamma)$, one takes $x_-(\tau)=0$;
with probability $1-\exp(-\gamma)$, one repeats the procedure of 4)
and calls the output $x_-(\tau)= \phi_{i_1}+...+\phi_{i_k}$.

\item 6) Repeat the steps 4) and 5) to generate two other variables $x_+'(\tau)$
and $x_-'(\tau)$.

\item 7) Compute $\chi(\tau)=e^{x_-(\tau)}/(e^{x_+(\tau)}-1)$,  $\chi'(\tau)=e^{x_-'(\tau)}/(e^{x_+'(\tau)}-1)$,
and $\phi_0(\tau)= \log[1+1/(\chi(\tau)+\chi'(\tau)+\chi(\tau)\chi'(\tau))]$.

\item 7) Replace one randomly chosen variable in the population, $\phi_\ell$,
by the new value $\phi_0(\tau)$.

\end{itemize}
The steps 3) to 7) must be repeated a large number of times, say $\cal T$.
One can compute the average of any function of $\phi$ as:
\begin{equation}
\int d\phi S(\phi) f(\phi) = \frac{4}{3 {\cal T}}
\sum_{\tau= {\cal T}/4+1}^{{\cal T}} f(\phi_0(\tau))
\end{equation}
(One forgets the first ${\cal T}/4$ iterations in order to 
reach a stationnary regime).
The integrals involving a variable $x$ distributed according to $B(x)$ can
be estimated by summing functions of $x_-$. For instance, the first term
in the zero temperature free energy (\ref{Phifinal}) is evaluated as 
\begin{equation}
-\int dx dz B(x) B(z) \log\left(e^x+e^z-1\right) = \frac{4}{3 {\cal T}}
\sum_{\tau= {\cal T}/4+1}^{{\cal T}}  \log\left(e^{x_-(\tau)}+e^{x_-(\tau)'}-1\right) \ ,
\end{equation}
and the two other terms are computed similarly, using the $x_-,x_-'$ values
from two (resp. three) successive time steps.

In practice, we used values of $N \sim 10000$ and 
${\cal T} \sim 1000 N$  which was enough to  reach 
the precision given below on the results.

For $\alpha<\alpha_d=3.921$, the algorithm converges towards the
solution $\phi_1=...=\phi_N=0$. This is the paramagnetic solution
where all the u-surveys are trivial.

For $\alpha>\alpha_d=3.921$, we find a new solution with a non-trivial distribution
$S(\phi)$. Computing the leading large $y$ behaviour of the 
zero temperature free energy function, $\Phi(y)\sim \Psi/y$,
on this solution using (\ref{Phifinal}), we find that $\Psi$ is negative for
$\alpha<\alpha_c=4.267$, while it is positive for $\alpha>\alpha_c$
(in the first version of this paper, we had reported the slightly
too small  value $\alpha_c=4.256$ quoted in\cite{MEPAZE}, because of a poor random number generator used in the population dynamics).

\subsection{Phase diagram of  the random  3sat problem}
The previous large $y$ analysis is in agreement with the direct
numerical iteration of the population dynamics of
sect.\ref{sect_popdyn}, but it allows to get a much more precise
determination of the thresholds.  The results of this section have
been obtained through the combined use of the numerical and analytical
method.

For $\alpha<\alpha_d$, the system is in the SAT phase, the solution is
paramagnetic, it is easy to find a solution.  Note that, although the
$t$ equation (\ref{EQ_t}) in principle allows for the existence of
nontrivial knowledges above $\alpha\simeq 1.64$, we have not found
such a solution and the only one which remains is the paramagnetic one
with $t=1$.

\begin{figure}
\centering
\includegraphics[width=12cm]{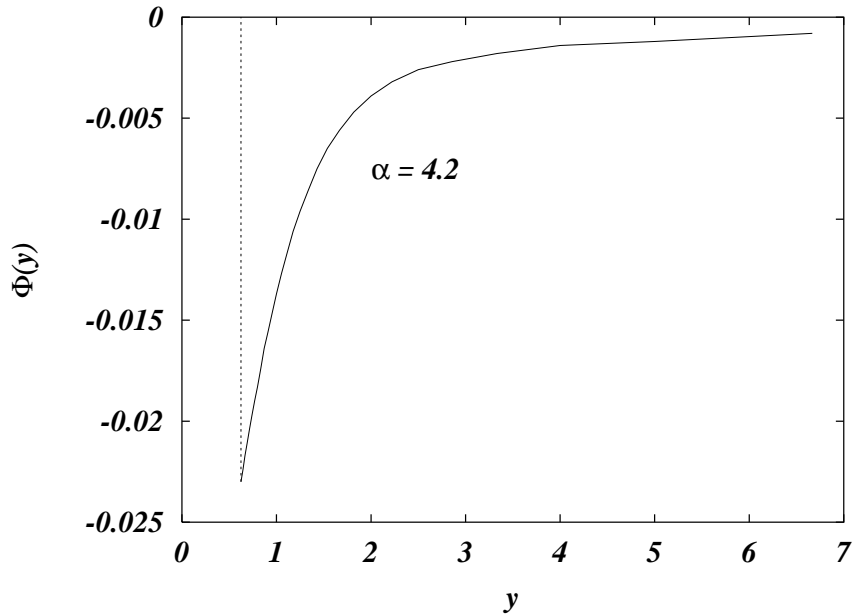}
\caption{Free energy $\Phi(y)$  versus the 
reweighting parameter $y$ for the random 3sat problem   at $\alpha=4.2$}
\label{free_random}
\end{figure}

For $\alpha_d<\alpha<\alpha_c$, we find a monotonously increasing $\Phi(y)$
function, which reaches its maximum $\Phi \to 0$ at $ y \to \infty$
(see fig. \ref{free_random}).

\begin{figure}
\centering
\includegraphics[width=12cm]{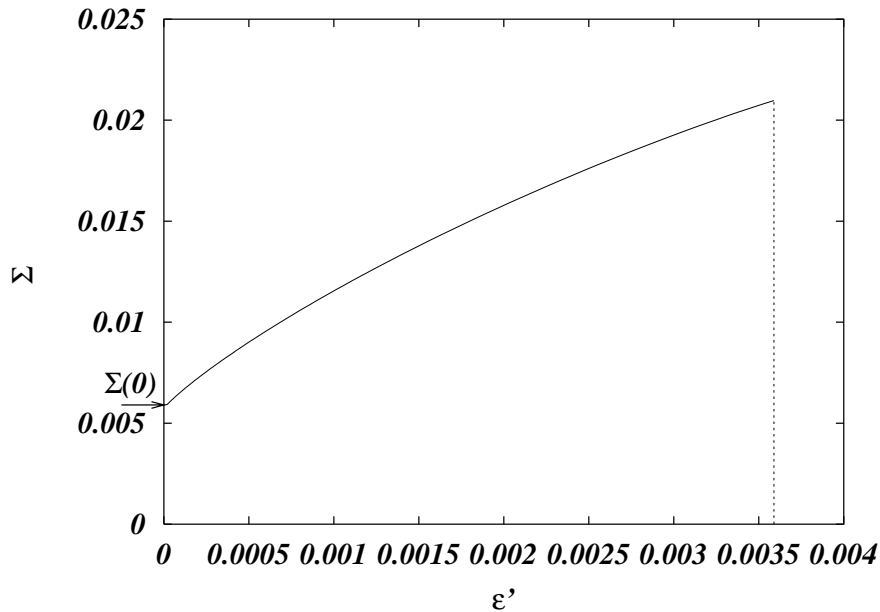}
\caption{Complexity $\Sigma$ versus the fraction of violated
clauses $\epsilon'=\epsilon/2$  for the random 3sat problem  at $\alpha=4.2$,
obtained from the Legendre transform of fig.\ref{free_random}}
\label{fig_complex_random}
\end{figure}

Fitting $\Phi(y)$ by a function $\sum_{\ell=1}^p \nu_\ell \exp(-y
\ell)/y$ with $p\in\{1,2,3\}$ gives a good and stable fit, from which
the Legendre transform (\ref{legendre}) is easily done. This allows to
reconstruct the complexity curve $\Sigma(\alpha,\epsilon)$ (see
fig.\ref{fig_complex_random}). It is found to be finite down to
$\epsilon=0$. This implies that there is an exponentially large (in $N$)
number of states with zero energy density.

\begin{figure}
\centering
\includegraphics[width=14cm]{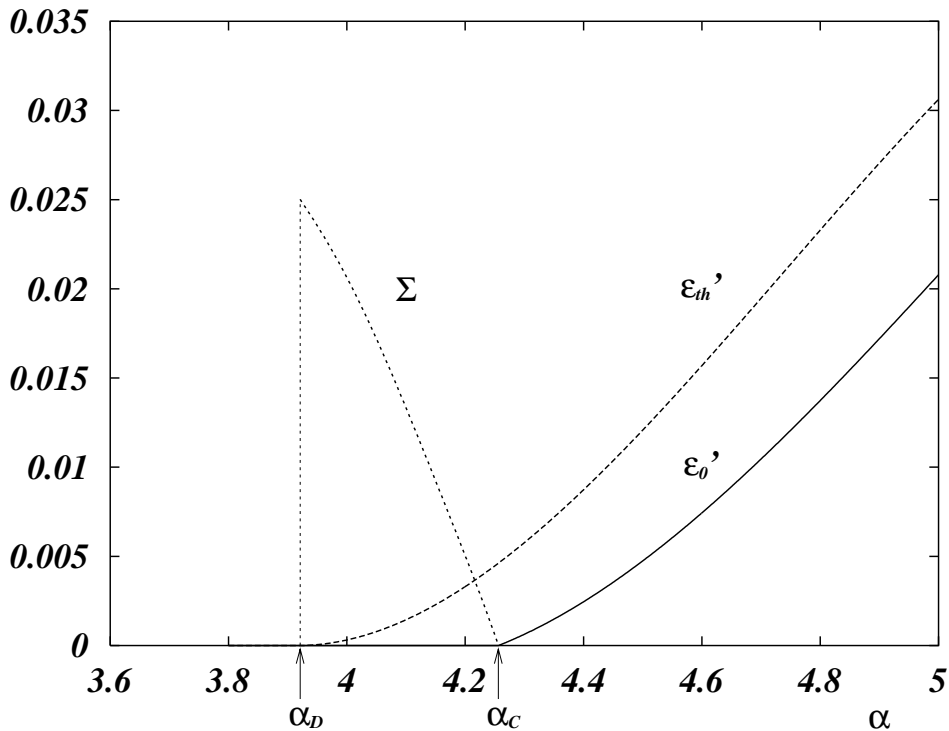}
\caption{ The phase diagram of the random 3sat problem. Plotted is
$\epsilon_0'=\epsilon_0/2$ (full line), the number of violated clauses
per variable, versus the control parameter $\alpha$ which is the
number of clauses per variable. The sat-unsat transition occurs at
$\alpha=\alpha_c \sim 4.267$.  The dashed line is
$\epsilon_{th}'=\epsilon_{th}/2$, the threshold energy (divided by
two) per variable, where local algorithms get trapped.  The dotted
line is the complexity $\Sigma$ of satisfiable states, equal to $1/N$
times the natural logarithm of their number.
}
\label{fig_phasediag}
\end{figure}

The complexity of these ground states is plotted in
fig.\ref{fig_phasediag}. We call this phase the Hard-SAT-Phase (HSP),
since in this regime the typical sample is SAT, but the proliferation
of states (most of which have strictly positive energy) makes it
difficult to find a solution.

For $\alpha_c<\alpha$, the function $\Phi(y)$ has a maximum at a
finite value $y^*$, and $\Phi(y^*)>0$. The complexity curve
$\Sigma(\alpha,\epsilon)$ starts at a positive energy density
$\epsilon_0=\Phi(y^*)$ (see fig.\ref{fig_complex_random}). This energy
density $\epsilon_0$ is the minimal number of violated clauses per
variable which will be found in almost all samples at large $N$.  It
is plotted in fig.\ref{fig_phasediag}. We are in the UNSAT phase.

This figure also shows the value of $\epsilon$ where the complexity
curve $\Sigma(\alpha,\epsilon)$ is maximum gives the energy where
there exists the largest number of states. This is the 'threshold'
energy density $e_{th}$ where a simple zero temperature Metropolis
algorithm (ZTMA) will be trapped. This implies that ZTMA should find
satisfying assignments only for $\alpha < \alpha_D$, in agreement with
the numerical results of \cite{Svenson}.  These predictions can be
tested most clearly through their generalization to single instances
which we discuss in the next section.

Previous statistical mechanics attempts at finding this phase diagram
culminated in powerful variational approximations using the replica
method, see ref. \cite{BMW} for the first results and
ref. \cite{FLRZ}, which predicted approximate values for the SAT/UNSAT
threshold -- $\alpha_c \sim 4.48$ in the case of ref.\cite{BMW} and
$\alpha_c \sim 4.39$ in \cite{FLRZ} -- with an intermediate phase
appearing above $\alpha_s \simeq 3.96$ \cite{BMW} and $\alpha_d \simeq
3.94$ with the Ansatz of \cite{FLRZ}.

The difference between the variational results and our cavity result
is both quantitative and qualitative: in ref. \cite{BMW} the predicted
nature of the intermediate phase is different with respect to ours
while in ref. \cite{FLRZ} the structure of the order parameter is
oversimplified.
In the present approach (as well as in \cite{FLRZ}) we work directly
at zero temperature ($T=0$), which has the advantage that we do not
need to study the subtle question of the limit $T \to 0$. The reason
why this limit is subtle is due to the fact that some of the local
fields, at low temperatures, vanish linearly in $T$, and thus
contribute to the local magnetization $m=\tanh (\beta H)$; we call
these fields evanescent fields. The local magnetization at $T=0$ is
zero for a zero field, it is equal to $1$ for a finite field, and it
takes an intermediate value $m \in ]-1,1[$ for an evanescent field.
The variational approach of \cite{BMW} focuses onto evanescent fields,
and finds a {\it continuous} phase transition at $\alpha_s \simeq
3.96$ where the evanescent fields in different states start to
cluster. However as these are all evanescent fields, this means that
the corresponding local magnetizations, in a given state, are not
frozen to $\pm 1$ but take some intermediate value, even in the $T \to
0$ limit.  In our $T=0$ cavity approach (as well as in \cite{FLRZ}),
the HSP corresponds to a {\it discontinuous} transition at zero
temperature, involving fields which are not evanescent, but are of
order one \cite{fields_comment}.  This means that, in a given state, a
finite number of local fields are non-zero integers, giving rise to
magnetizations $\pm 1$, as one could expect at zero temperature.  This
phenomenology cannot be found by considering evanescent fields. Its
study with replicas would require using a more complicated Ansatz.

Note that our approach working directly at $T=0$ also has its
limitations, for instance we are unable to determine precisely the
self overlap (or the typical radius) of a state, or its internal
entropy, precisely because we do not control the evanescent fields.
We leave for future work the delicate study of the small $T$ region of
the phase diagram. Let us just notice here that a population dynamics
study of this region with the 1RSB finite temperature Ansatz of
\cite{MP_Bethe} shows that the distributions of local fields tend to
peak on integers when the temperature goes to zero in the HSP
\cite{GP_pop}. This is a strong argument in favor of the exactness of
this 1RSB solution (i.e. the fact that we do not need to go to higher
order RSB), as argued in \cite{MP_T0}.

\section{Survey propagation: configuration space analysis on a single
instance}
\label{sect_SP}

The analysis of the iterative equations for the probability
distributions of messages of the previous sections was done by using a
population dynamics algorithm, which performs an average over the
underlying random factor-graphs.  At each step of the iteration, a
random choice of coupling constants as well as neighborings nodes is
performed with the proper probability distribution. While such an
averaging step is central if one wants to estimate typical properties,
the iterative equations make perfect sense on a single specific
instance. The order parameters arising from the cavity equations,
namely the u-surveys, are histograms of probability distributions of
cavity biases. They determine the bias of each spin in all metastable
states of a given energy density for a given instance of the
underlying factor-graph.  This is a very important piece of
information which can be exploited to study specific problems and to
invent new algorithms.

In the large N limit, we expect that the cavity assumptions hold for
locally tree-like factor graphs and we may use u-survey propagation to
have access to the properties of optimal states of minimum
energy. Here we shall develop one such application for random 3-sat
which is a concrete world-wide benchmark for search algorithms.
However the idea of exploiting the information on optimal states
carried by the functional order parameter is rather general and we
expect algorithmic applications in different fields.

Whenever the factor graph representing the problem does not lead to
clustering within states, in practice whenever loops are short enough,
one should think to the present approach as a first step of a sequence
of possible approximations. 

As is well known, the so called Cluster Variation Method\cite{CVM}
provides a systematic scheme that can be adopted to improve the
approximate results given by the cavity approach\cite{YedFreeWei}. The latter
corresponds to the so called Bethe approximation which is the first
step in the cluster variation scheme. Our present approach deals the
Bethe approximation in a frustrated case.  While a great deal of work
has been done concerning higher order cluster approximations for
simple models, the corresponding analysis for frustrated systems such
as spin-glasses or hard-combinatorial problems over non locally
tree-like graphs is largely unexplored.

\subsection{The Survey Propagation (SP) algorithm}
\label{SingleSample}

In the ordinary sum-product algorithms~\cite{factor_graph} as described
in sect.  \ref{sect_sum_prod}, the messages arriving at a node are
added up and then sent to a function node. Next, the function node
transforms all input signals into a new message which is sent to the
descendant variable node. At each time step, on the links of the
factor graph there are signals traveling, just like in a communication
network. SP works with the same principle, but now the messages
traveling along the links of the factor graph are u-surveys of usual
messages over the various possible states of the system at a given
value of the energy (or rather, in practice, at a given value of the
reweighting parameter $y$). Of course this higher level of description
is useful only when there are many states, which will be typically the
case in hard optimization problems. One a-priori drawback of the
approach is that the messages are complicated, being functional
probability distributions: a cavity-bias is already a parametrization
of a probability distribution (which turns out to be parametrized by a
single variable in our case of binary spins, but could be more
complicated in general); Here the u-surveys are probability
distribution functions of these cavity-biases. In cases like Ksat at
$T=0$ in which the standard messages can take only few values (say
$r$), a u-survey is given by the probabilities of these values,
i.e. by $r-1$ real numbers and the SP can be implemented easily.  This
is one big advantage of working directly at $T=0$, but we believe that
the SP method could also be used more generally at finite temperature
or with continuous variables, by using well adapted parametrizations
of the cavity-biases.

SP is defined for one given value of the reweighting parameter $y$ and
one given instance, with $N$ variable nodes and $M$ function
nodes. Its basic ingredients are the u-surveys.  Each edge $a \to j$
from a function node to a variable node $j$ carries a u-survey $Q_{a
\to j}(u)$. The algorithm finds these u-surveys by a message passing
procedure detailed below, and finds simultaneously all the h-surveys
$P_{i \to a}(h)$.  Once these are known, one can compute the so called
local field distributions and the 'zero temperature free energy' for
this instance .  The local field distribution $P_i(H)$ on a variable
node $i$ is the distribution, over all states selected by the
reweigthing parameter $y$, of the total local field $H$ acting on spin
$\sigma_i$ (see (\ref{Hdef})).  It is given by
\begin{equation}
P_{i} (H)=C_i \int \prod_{a \in V(i)} du_a
Q_{a \to i} (u_a)  \delta \left( H -\sum_{a \in V(i)}
u_a  \right) \exp\left(y \vert \sum_{a \in V(i)} u_a \vert \right)
\label{local_field}
\end{equation}
$C_i$ being the normalization constant.

We show in appendix B that the zero temperature free energy $\Phi(y)$
density of {\it this sample} can be computed as a sum of contributions
$\Phi^f_{a}(y)$ for each function node $a$, corrected by the
contributions $ \Phi^v_i(y)$ from each variable node $i$, weighted by
a factor $n_i-1$, where $n_i$ is the connectivity of variable node
$i$:
\begin{equation}
\Phi(y)= \frac{1}{N}\left(\sum_{a=1}^M \Phi^f_{a}(y)-\sum_{i=1}^N \Phi^v_i(y) (n_i-1)\right)  \ ,
\label{free_onesamp1}
\end{equation}
where:
\begin{eqnarray}
\Phi^f_{a}(y)&=&-\frac{1}{y} 
\log \left\{
\int \prod_{i \in V(a)}
\left[ \prod_{b\in V(i)-a} Q_{b\to i}(u_{b\to i}) d u_{b\to i} \right] \right.
\nonumber
\\ 
\nonumber
&&\left.
\exp \left[ -y \min_{\{ \sigma_i,i \in V(a)\} } \left( E_a -\sum_{i \in V(a)} \left[
\sum_{b\in V(i)-a}u_{b\to i} \right] \sigma_i \right) \right] \right\}
 \nonumber \\
\Phi^v_i(y)&=&-\frac{1}{y} \log \left \{ 
\int \prod_{a \in V(i)} du_a
Q_{a \to i} (u_a)  \exp\left(y \vert \sum_{a \in V(i)} u_a \vert \right)
 \right \}=-\frac{1}{y} \log (C_i)
\label{free_onesamp2}
\end{eqnarray}
The form (\ref{free_onesamp1}) is the familiar one for the free energy
in the Bethe approximation \cite{Yedidia}, and indeed one gets back the
usual result using (\ref{free_onesamp2}) in the $y \to 0$ limit.  The
generalization to $y \ne 0$ given in (\ref{free_onesamp2}) adds the
effect of the reweighting terms due to level crossings.  The origin of
these terms is exactly the same as in sect. \ref{sect_cavity}. 

Let us now explain how SP works. We start with a general presentation
of the algorithm, which applies to any optimization problem
involving binary variables, characterized by a given factor
graph. Some details of the implementation for the 3sat problem will be
given below.

\begin{figure}
\centering
\includegraphics[width=12cm]{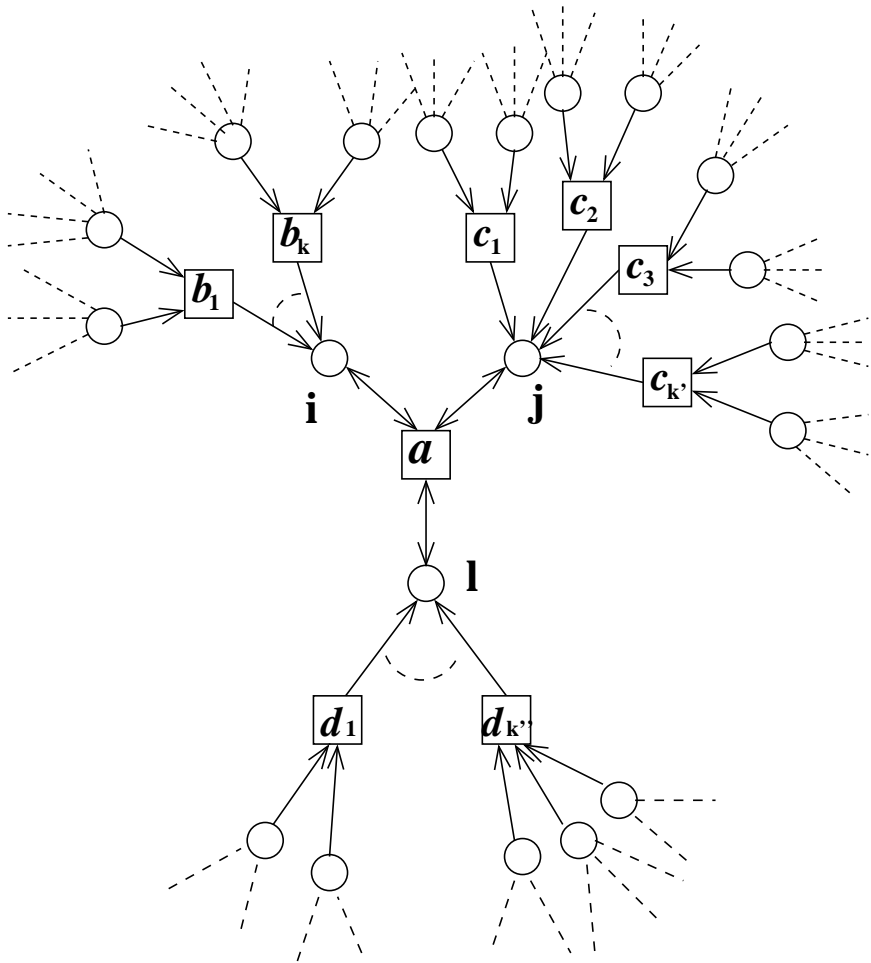}
\caption{Function node a and its neighboring graph}
\label{node}
\end{figure}

\begin{itemize}
\item[(0)] All the u-surveys $Q_{a \to i}(u)$ are initialized
randomly.

\item[(1)] Function nodes are selected sequentially at random; for
each such node $a$, we update the u-surveys as follows (see
fig. (\ref{node})):

\begin{itemize}
\item[(1.1)] for each variable node $i$ connected to the selected
function node $a$, we compute the h-survey $P_{i \to a}(h)$ as a reweighted
convolution, see fig. (\ref{node}),
\begin{equation}
P_{i \to a}(h )= C_{i \to a} \int du_1...du_{k} 
Q_{b_1 \to i} (u_1)...Q_{b_{k} \to i}(u_{k}) 
\delta \left(h-\sum_{a=1}^{k} u_a \right) \exp \left( y\vert \sum_{a=1}^ku_a\vert \right)
\label{P1siter}
\end{equation}
\item[(1.2)] successively, the u-surveys on all edges $a \to i$
connected to $a$ are updated using these h-surveys:
\begin{equation}
Q_{a \to i}(u)=C_{a \to i} \int dg dh P_{j \to a}(g) P_{\ell \to a}(h)
 \delta\left( u-\hat u_J(g,h)\right) \exp \left( y \left[\hat w_J(g,h)
 -\vert g\vert-\vert h \vert\right] \right)
\label{Q1siter}
\end{equation}
\end{itemize}
($C_{i \to a},C_{a \to i}$ are normalization constants).

\item[(2)] The iterative process of step (1) continues until
convergence is reached. If the process converges, the corresponding
stable set of u-surveys are used to compute the $N$ local field
distributions using (\ref{local_field}), and the zero temperature free
energy $\Phi(y)$ given in (\ref{free_onesamp1},\ref{free_onesamp2}).

\end{itemize}

The above procedure can be repeated for different values of the reweighting
$y$ so that the complexity $\Sigma(y)=\partial \Phi(y) / \partial(1/y)$ and
the energy density $\epsilon(y)=\partial (y \Phi(y)) / \partial y$ of states can
be estimated. The parametric plot of $\Sigma(y)$ versus $\epsilon(y)$, varying
$y$, gives the complexity $\Sigma(\epsilon)$ of states of energy $E=N\epsilon$.

When it converges, SP allows to get an order parameter (the set of all
the surveys), a zero temperature free energy density $\Phi(y)$, and a
complexity curve $\Sigma(\epsilon)$ for one given instance.  What is the
meaning of these quantities in general is an open question. A given
instance has a finite value of $N$ , and therefore the notion of
'state' is not easy to define. Roughly speaking one can think that for
large $N$, there might exist some effective 'finite $N$ states', such
that the number of spins to flip in order to reach one state from
another one is large, leading to a separation of scales of the number
of spins involved between the intra-state moves and the inter-states
moves. Such a situation would generally be difficult to handle for
search algorithms, and this is where SP could be quite useful.  In
order to get a first understanding of these questions, we have
experimented SP on single instances of the random 3sat problem for
large values of $N$.

\subsection{The case of random 3sat}
We consider one instance of the 3sat problem, chosen randomly as in
sect.  \ref{sect_rangraph}, with energy
\begin{equation}
E=2\sum_{a = 1 } ^M \prod_{j \in {V(a)}}\frac{1+J_{a \to j}
\sigma_j}{2} \ .
\end{equation}
For 3sat the cavity-biases on a given link $a \to j$ takes values
$u_{a\to j} =0,-J_{a \to j} $; The corresponding survey is $Q_{a \to
j}(u)=c_{a \to j} \delta(u)+(1-c_{a \to j}) \delta(u-J_{a \to j})$.
The full set of u-surveys is characterized by the $3 M$ numbers $c_{a
\to j}$ which are updated according to the SP algorithm described
above, until convergence.  The results of our numerical experiments
are the following.

\subsubsection{ The paramagnetic phase}

For $\alpha <\alpha_d$, SP converges toward the trivial paramagnetic
solution $Q_{a \to i}(u)=\delta(u)$, for all $a \to i$ edges. Local
field distributions are also trivial, $P_i(H)=\delta(H)$ $\forall i$,
and no information can be gained on the fine structure of ground
states. In this region, there is a single state and the statistical
properties of the zero energy configurations are totally driven by its
entropy. A different formulation of the cavity approach, in which the
proper $\beta \to \infty$ limit is taken and the evanescent %M
fields are computed, could reveal some finer
information for this paramagnetic phase, which however is known to be
trivial %M\cite{note_0} 
from the algorithmic point of view.

\subsubsection{The intermediate phase}

For $\alpha_c>\alpha >\alpha_d$, that is in the glassy region, the
random sequential updating of the iterative process converges to a
unique non-trivial solution, provided $y$ is large enough.  In
practice, we start from $y$ large, like e.g. $y=6$ (remember that
the corrections to the $y \to \infty$ limit are exponentially small),
run SP, and after finding a solution for the u-surveys we decrease $y$
(e.g. $y \to y-.2$) and rerun SP using the previous u-surveys as a
starting configuration for this new $y$. This speeds up the
convergence. Below some value of $y$ the non-trivial solution
disappears abruptly and the algorithm converges to the paramagnetic
solution.

\begin{figure}
\centering
\includegraphics[width=12cm]{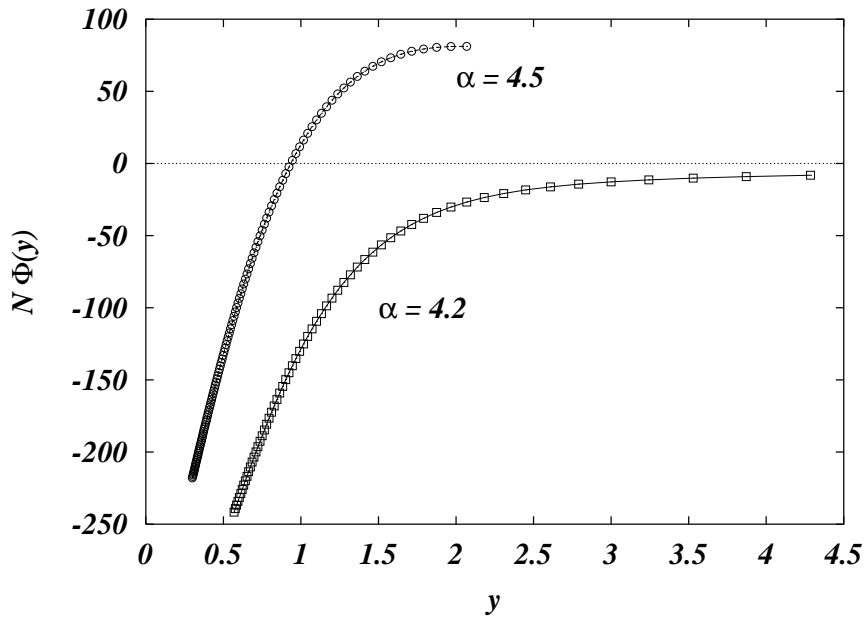}
\caption{Extensive zero temperature free energy $N \Phi(y)$ versus
reweighting $y$ for two specific instances of size $N=10000$ with
$M=42000$ and $M=45000$ clauses respectively. ($\alpha=4.2$ and
$\alpha=4.5$).}
\label{free}
\end{figure}

In this region of $\alpha$, the solution space as well as the
configurations of higher energy become divided into an exponential
number of states.  To compute the complexity, we measure the free
energy $\Phi(y)$, see fig. (\ref{free}), and we perform the Legendre
transform numerically.

\begin{figure}
\centering
\includegraphics[width=12cm]{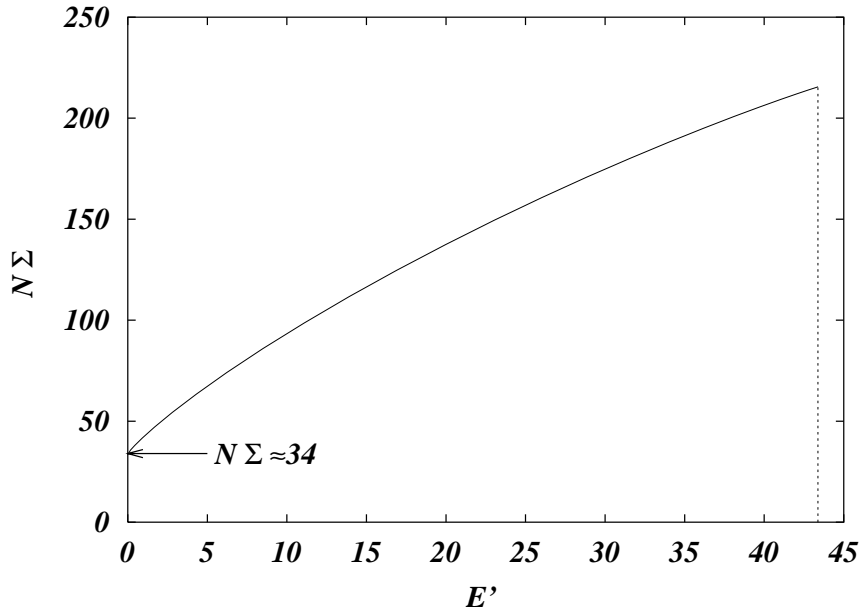}
\caption{Extensive complexity, $N \Sigma$, versus the total number of
violated clauses $N E'$ ($=N E/2$), for the specific instance of size
$N=10000$, $M=42000$ studied in fig.\ref{free}. The complexity is
obtained as the Legendre transform of the zero temperature free
energy. }
\label{complexity}
\end{figure}

The curve of the total complexity $N \Sigma$  versus the 
total energy $N \epsilon$  for one sample of random
3-sat with $N=10000$ and $M=42000$ is given in
fig. (\ref{complexity}). One finds $N \Sigma(E=0) \sim 34$, meaning that
the zero energy (SAT) states are predicted to be exponentially
numerous, $e^{34}$ at the leading exponential order (remember that
each such state itself contains a large number of spin
configurations~\cite{MoZe_prl}).  The threshold states have an energy of
approximatively $44$ violated clauses and their number is predicted to
be  about $e^{216}$.  A cross check of such predictions is given by
the behaviour of ZTMA which cannot cross energetic barriers.  It can
be shown that for random 3-sat zero energy moves allow to explore
configurations within each state and therefore we expect such
algorithms to get trapped in the most numerous ones (the threshold
states).  Indeed, extensive numerical simulations of ZTMA on many
samples of different size (ranging from few hundreds to $10^5$) and
for different values of $\alpha$ confirm such scenario.  As a
representative example, we report that for the sample whose compexity
is plotted in fig. (\ref{complexity}), repeated runs of ZTMA get stuck
at an energy sharply peaked around $48$ violated clauses, with a small
residual dependence of the final energy on the simulation time (the
final energy found by ZTMA shows a power law behavior on the total
number of spin flips).

We have checked that the functional order parameter given by the
u-surveys and the local field distributions carry precise information
concerning the space of solutions for one given sample. Working with
$N$ not too large, some SAT configurations can be found efficiently by
good algorithms like e.g.  walksat-35 \cite{walksat,SATLIB}. We have
collected a large set (1000) of uncorrelated SAT configurations by
running this algorithm many times with random initial conditions.  In
each such 
configuration $\omega$, the spin $\sigma_i$ takes a value
$\sigma_i^\omega$, and we have computed, for each given site $i$, the
average of the whole set SAT configurations $\omega$.  Next we have
compared the above results with the predictions of SP as follows.

\begin{figure}
\centering
\includegraphics[width=12cm]{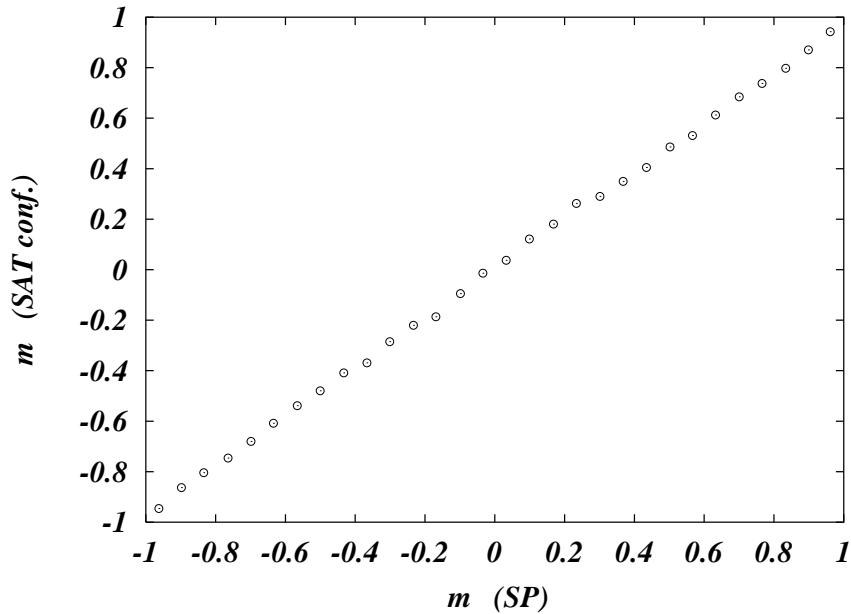}
\caption{The bias of the variables predicted by SP (with $y=8$)
compared with the one measured analyzing SAT configurations from the
same sample ($N=10000$,$M=42000$)}
\label{mk}
\end{figure}

We first have selected the states with minimal energy, by picking the
value of $y$ which maximizes $F(y)$. Here this is $y = \infty$ and for
practical computation it was enough to choose $y$ sufficiently large
(corrections are exponentially small).  According to
eq.(\ref{local_field}), the field $H_i=\sum_a u_{a\to i}$ in each
state is an integer valued variable which can be computed from the
u-surveys. The total weights
\begin{equation}
 w_i^+=\int_{0^+}^\infty dH P_i(H) \ \ \ ; \ \ 
w_i^-=\int_{-\infty}^{0^-} dH P_i(H)
\label{wipmdef}
\end{equation}
of $P_i(H)$ on positive (resp. negative) integers give the fractions
of zero energy states where $\sigma_i$ is fixed to $1$ (resp. to
$-1$).  As displayed in fig. (\ref{mk}), we find a remarkable
agreement between the local magnetizations $w_i^+-w_i^-$ predicted by
SP and the local magnetizations measured by averaging over the ground
states found by the walksat algorithm. In the figure we report data
for $N=10000$ and $M=42000$: we have devided the local magnetization
in 30 intervals and labeled spins according to the prediction of
SP. Next on such a partitioning of spins we have taken the average
over the configurations found by walksat. (The remarkable agreement
of numerical and SP results indirectly shows that walksat is a good
uniform sampler.)

The weight in $H=0$ of $P_i(H)$, $w_i^0=1-w_i^+-w_i^-$, measures the
tendency of a variable to be under constrained: for instance,
variables which belong to very few clauses have $w_i^0=1$.

\subsubsection{The UNSAT phase}

For $\alpha >\alpha_c$, SP predicts a positive ground state energy
with zero complexity, whereas excited states remain exponentially
numerous. Proper tuning of the reweighting, that is choosing $y$ so
that the complexity vanishes, allows to predict the ground state
energy and to evaluate the probability distribution of effective
fields for each variable. In this regime SP is found to converge only
when the reweighting parameter is well chosen. For small values of
$y$, SP converges to the paramagnetic solution or to the RS
solution. For intermediate values of $y$, SP converges to the
non-trivial solution whereas for larger values of $y$, SP stops
converging. The range of $y$ values for which SP converges to the
non-trivial solution is sufficient to determine the free energy. An
example is given in fig.\ref{free}.

For large values of $\alpha$, we expect multiple nested clustering
phenomena to appear, that is continuous replica symmetry
breaking\cite{leuzzi}.  This scenario could be analyzed by a further
generalization of SP which is beyond the scope of this work.

\section{Survey Propagation as a source of new algorithms 
for hard optimization problems}
\label{sect_SID}

The previous section has shown how SP can give rather precise answers
on the structure of the space of configurations and the ground state
energy of the random 3sat problem. Here we shall stay within this
problem and ask the following natural question:

\begin{quote}
{\bf Given a random 3-sat formula of size N, how can we take advantage of
SP in order  to find optimal configurations?}
\end{quote}

If SP could predict with very high accuracy the value of the GS energy
of a given formula, it could also predict its satisfiability.  Then
one could proceed in finding a satisfying assignment just by
converting the decision algorithm into a search algorithm as
follows~\cite{Garey_Johnson}.  A variable is selected and fixed to one
value.  We then use SP to evaluate the GS energy of the subproblem of
size $N-1$ and decide whether it is still SAT or not. If the
subproblem is SAT then we keep the assignment, otherwise the opposite
value of the binary variable is chosen. The process is repeated untill
all the variables have been exhausted (in at most $2 N$ steps).  If
along this reduction process the subsystem becomes a paramagnet, then
SP becomes ineffective and another search algorithm must be used on
the subsystem (but for a paramagnet it is very easy to find the ground
state).

The above scheme, however, suffers from finite size effects and from
the imprecision in the determination of the ground state energy, a
fact which is particularly important close to $\alpha_c$. Moreover it
does not take advantage of the information provided by the u-surveys.

\subsection{Categories of  variables in one specific  instances}

A somewhat coarse grained information contained in the u-surveys, once SP has
reached convergence, is the one given by the total weights $w_i^{\pm}$ of the
local field distribution which gives the fraction of states where the spin
$\sigma_i$ is positive (resp. negative). Having computed these weights, we may
distinguish three reference types of spins (of course all the intermediate
cases will also be present): the {\it paramagnetic} ones with $w_i^0 \sim 1$,
the {\it biased} ones with $w_i^+ \sim 1$ or $w_i^- \sim 1$ and the {\it
balanced} ones with $w_i^+ \simeq w_i^-$ and $w_i^0$ small.

\begin{figure}
\centering
\includegraphics[width=12cm]{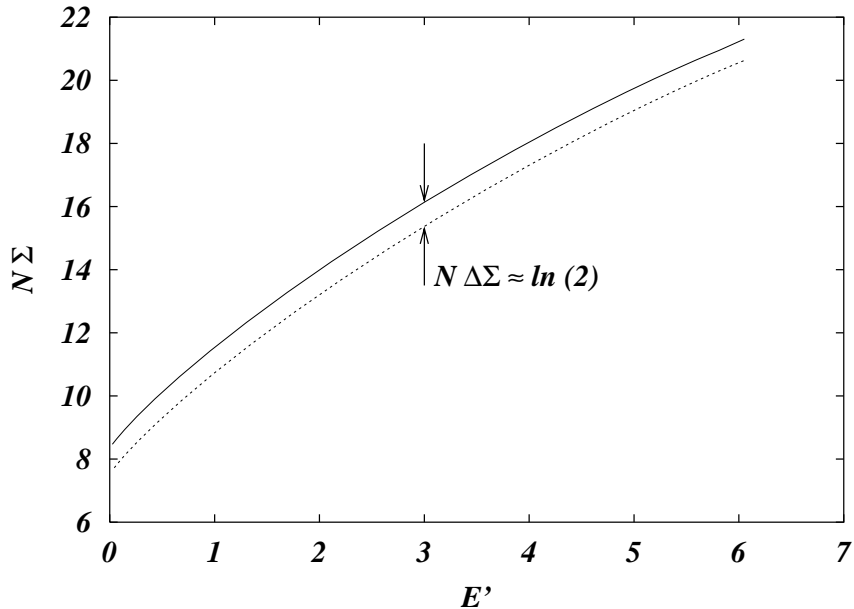}
\caption{Effect of fixing a single balanced spin on the complete
complexity curve $N \Sigma$ versus the number of violated clauses (
i.e. one half of the extensive energy) of an instance of 3sat with
$N=1000$ and $M=4200$. The difference in the two curves is very close
to $\ln 2$.}
\label{single}
\end{figure}

In order to characterize the differences between these various types
of variables, we have performed a few numerical experiments and
analyzed the effect of fixing one such spin on the structure of
states of the subproblem of size $N-1$.  As displayed in the
complexity curve of fig.(\ref{single}),
the three types of spins produce different effects, consistently with
the interpretation of the order parameter.  Fixing a biased spin does
not alter the structure of the states and the complexity changes
smoothly.  Fixing a paramagnetic spin has an effect only on the
internal entropy of the states (which we cannot measure) but leaves
the energy unaltered.  Interestingly enough, balanced spin have an
enormous effect: the most balanced ones produce a decrease very close
to $\ln 2$ in the complexity, indeed half of the states are eliminated
by fixing one single balanced variable!.

\subsection{Survey inspired decimation (SID) algorithm}

One strategy for using this information in order to produce an
optimization algorithm is to fix as many variables as possible without
altering the ground state energy, evaluated step by step as the size
of the problem decreases. Eventually, either all variables have been
fixed or (more likely) the remaining variables turn out to be
paramagnetic (i.e. $P_i(H)=\delta(H)$, $\forall i$), in which case a
simple search process can be run to find the complete ground state
configuration.

A straightforward implementation of the above ideas provides a simple
algorithm that can be used to find solutions to random 3sat in the hard
region $\alpha \in [\alpha_d,\alpha_c]$.  We do not expect this
implementation to be the most efficient one, in that no particular
strategy has been worked out to optimize the decimation process. The
scope of this first implementation consists in showing the
potentiality of the novel algorithmic scheme and we leave for future
work the design of optimized versions of the algorithm or applications
in different contexts.

The overall idea underlying the search process is rather simple. At
each time step a single variable is fixed according to the outcome of
SP and the effect of such fixing is used to simplify the problem. The
size of the problem reduces from $N_t$ to $N_t-1-S_t$, where $S_t$ is
the number of variables which become fixed due to the simplification
of the problem: satisfied clauses are eliminated, unsatisfied K
clauses are transformed into (K-1) clauses. $K=1$ clauses need to be
satisfied and therefore their variables are properly fixed (unit
clause propagation) leading to further spin elimination.

At the beginning of the process, randomly chosen balanced spins can be
fixed in order to reduce the number of states. At each step one may
compute the free energy to detect the onset of violated clauses. One
may also evaluate the function $\Phi(y)$ to have an estimate of the
complexity. Successively, biased spins are fixed.  Whenever a
paramagnetic state is found, or at some intermediate steps, a rapid
search process like simulated annealing at a fixed cooling rate or
walksat is run on the sub-system.  We may end up either by having
found a solution or by having still few violated clauses. In the
latter case we may simply restart. The sketch of the SID algorithm is:

{\tt

\begin{itemize}
\item[0.] Random initial condition for the cavity-biases

\item[1.] Run SP and evaluate $\{P_i(H)\}$, or $\{w_i^{+}, w_i^{-},
w_i^{0}\}$, and $\Phi(y)$.

\item[2.] Check for a paramagnetic state and in case (or at some
intermediate step) run a fast local search process (e.g. simulated
annealing or walksat). If a solution is found output ``SAT'' and stop.

\item[3.] Select and fix the most biased variable (the one with the
largest $\vert w_i^{+}- w_i^{-}\vert$) and simplify the problem.

\item[4.] If the problem is solved completely by unit clause
propagation, then output ``SAT'' and stop.  If no contradiction is
found then continue the decimation process on the smaller problem
({\bf go to 1.})  else (if a contradiction is reached) restart ({\bf
go to 0.})

\end{itemize}

}

Extensive numerical experiments on random 3sat instances at
$\alpha=4.2$ with size up to $N=10^5$ have shown a remarkable
efficiency of SID. While the process of fixing a single variable takes
some time ($O(N)$ operations) the number of 
assignments explored is very small. At $\alpha=4.2$ typically a single
run of SID (i.e. with no restarts) leads to a solution.  Closer to the
critical $\alpha$, few restarts might be necessary in order to find a
configuration of strictly zero energy. However, at each run the
typical energy found by SID is very close to zero, well below the
energy at which simulated annealing gets stuck. A detailed description
of the numerical experiments will be given in a furthcoming paper
\cite{BrMeZe}. We just mention that the largest public benchmarks of
random 3sat \cite{SATLIB} have been solved efficiently by SID.

\begin{figure}
\centering
\includegraphics[width=12cm]{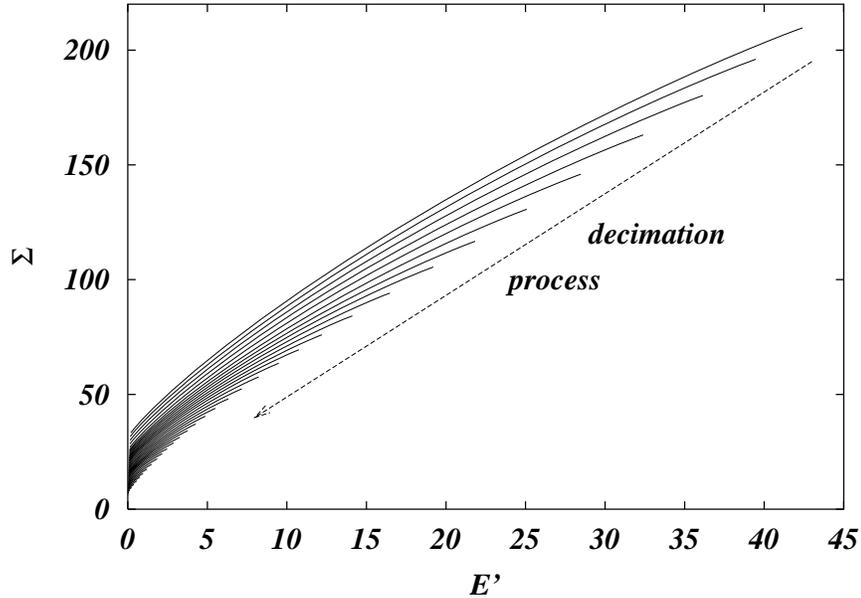}
\caption{ Evolution of the complexity curve upon decimation: The
$\Sigma$ versus $E'=E/2$ curve is shown when $200,400,600,...$ spins
have been fixed according to the SID algorithm ($N=10000$,$M=42000$
for this sample).}
\label{sigma_decimation4p20}
\end{figure}

In fig. (\ref{sigma_decimation4p20}) we show the evolution of the
complexity under SID. For a sample of size $N=10000$ at $\alpha=4.2$
we evaluate the complexity curve every 200 decimation steps until a
paramegnetic state is reached. SID acts by eliminating clusters of
solutions and hence reducing the complexity of the ground state down to
the point where very few clusters remain.

\section{Conclusion}
\label{sect_concl}

We have derived here two main results. The first one concerns the
phase diagram of the random Ksat problem, and establishes the
existence of an intermediate phase where the problem is SAT but the
solution is difficult to find because of the existence of many
states. We would like to point out that the cavity method which we
have used here is not rigorous: it relies on some hypotheses which can
be true only for large systems and are thus difficult to prove
(although this can be done in some cases \cite{talag}). However
the experience gained from similar problems, together with numerical
results of this and previous papers, indicates that this solution is
likely to be the correct one. If higher order replica symmetry
breaking effects would show up, one can believe that in any case their
quantitative influence on the results should be rather small. It should
also be noted that the very same cavity strategy which we have used
here has been tested on a variant of the Ksat, the XORsat problem,
which can be also solved by exact methods \cite{MRZ,DubCocMon}. In
this case, the existence of the intermediate phase has been confirmed
and all the predictions (qualitative and quantitative) from the cavity
method have been checked rigorously \cite{MRZ}.

The second result gives a new class of message passing algorithms for
solving optimization problems in the regime where the proliferation of
metastable states slows down a lot all the local algorithms. One such
algorithm turns out to be quite powerful on the case of random 3sat
problems. Clearly a lot of work needs to be done in order to develop
such algorithms in various contexts and test them against traditional
strategies. A more direct derivation and understanding of the
algorithm, and in particular of the nontrivial reweighting term, would
also be welcome.

Finally, we expect that the {\it single sample SP method} at finite
temperature can become a useful tool in analyzing the fine structure
of order parameters in disordered systems and other complex
systems.

\section*{Acknowledgements}
We thank G.Parisi, B. Selman and J.S.Yedidia for useful discussions
and encouragement, and S. Mertens for pointing out the
inaccuracy in our first estimate  of $\alpha_c$.  
RZ thanks the LPTMS, Orsay, for the kind
hospitality.

\appendix
\section{Large $y$ expansion for the 3-sat problem}
We give here some details on the solution of the population
dynamics equation (\ref{QRSBequation_3sat}) of the 3sat problem
at large $y$. 
We start from the self-consistent equation (\ref{EQ_RHO1}) for the
distribution of the rescaled weights of the u-surveys in $u=1$,
and we recall that $f_q^{(m,n)}({ \bf \eta} )$ is defined as the probability
that $u_1+...+u_k=q$, given that $Q_1(u_1),...,Q_m(u_m)$ are of type $b_+$,
and $Q_{m+1}(u_{m+1}),...,Q_{m+n}(u_{m+n})$ are of type $b_-$.
The quantities $A_0,B_0$ are expressed in terms of three
numbers $g_+,g_-,g_0$:
\begin{equation}
g_+({\bf \eta }) \equiv\sum_{q=1}^k f_q^{(m,n)}
( {\bf \eta} ) e^{y |q|}
\ \ ; \ \ 
g_-({\bf \eta })\equiv\sum_{q=-k}^{-1} f_q^{(m,n)}
({ \bf \eta} ) e^{y |q|} \ \ ; \ \
g_0({\bf \eta })\equiv f_0^{(m,n)}( {\bf \eta} ) \label{def_g1}
\end{equation}
as:
\begin{eqnarray}
A_0&\equiv& g_+({\bf \eta}) g_+({\bf \eta'}) \nonumber \\
B_0& \equiv&  g_+({\bf \eta})[ g_-({\bf \eta'})+g_0({\bf \eta'})]+
 g_+({\bf \eta'})[ g_-({\bf \eta})+g_0({\bf \eta})]+ [ g_-({\bf \eta'})+g_0({\bf
\eta'})] [ g_-({\bf \eta})+g_0({\bf \eta})]
\end{eqnarray}

The next step consists in introducing the joint probability
distribution $P^{(m,n)}(g_+,g_-,g_0)$
\begin{eqnarray}
P^{(m,n)}(g_+,g_-,g_0) &\equiv& \int d\eta_1...d\eta_{m+n} 
\rho(\eta_1)...\rho(\eta_{m+n})
\delta\left( g_+ - \sum_{q=1}^k f_q^{(m,n)}({\bf \eta}) e^{y |q|}\right)
\nonumber \\
&\times&
\delta\left( g_- - \sum_{q=-k}^{-1} f_q^{(m,n)}({\bf \eta}) e^{y |q|}\right)
\delta\left( g_0 - f_0^{(m,n)}({\bf \eta})\right) \ .
\end{eqnarray}
 Eq. (\ref{EQ_RHO1}) reads
\begin{eqnarray}
\rho(w_0) &=& \frac{1}{1-t} \sum_{k=0}^\infty \sum_{m=1}^{k} 
\sum_{n=0}^{k-m} C_{k,m,n} \sum_{k'=0}^\infty \sum_{m'=1}^{k'}
\sum_{n'=0}^{k'-m'} C_{k',m',n'} \int dg_+ dg_- dg_0 dg_+' dg_-' dg_0' \nonumber
\\ & &
P^{(m,n)}(g_+,g_-,g_0) P^{(m',n')}(g_+',g_-',g_0')
\delta\left(w_0-\frac{A}{A+B}\right)
\label{EQ_RHO}
\end{eqnarray}
where the coefficients $C_{k,m,n}$ are given in (\ref{Ckmn_def}), and
$A,B$ are given in (\ref{ABdef}).

The scaling of the  weights of the u-surveys in $u=1$ $w_0$ with $e^{-y}$ 
at large $y$ is  consistent with the self-consistency equation (\ref{EQ_RHO1}).
In this limit we find $g_+,g_-,g_0 \sim O(1)$, and these quantities
simplify to (up to corrections of order $e^{-y}$):
\begin{eqnarray}
g_+({\bf \eta}) &=&  (1+\eta_1) ...  (1+ \eta_m)-1
\nonumber \\
g_-({\bf \eta})&=&(1+\eta_{m+1}) ...  (1+\eta_{m+n})-1 \nonumber \\
g_0({\bf \eta})&=& 1 \label{def_g2}
\end{eqnarray}
Therefore, the three variables $g_+,g_-,g_0$ become uncorrelated
random variables in the large $y$ limit, with a distribution:
\begin{equation}
P^{(m,n)}(g_+,g_-,g_0) = \Omega^{(m)}(g_+) \Omega^{(n)}(g_-)\delta(g_0-1) \ ,
\label{def_Pmn}
\end{equation}
where
\begin{equation}
 \Omega^{(\ell)}(g)=\int d\eta_1...d\eta_\ell \hat \rho(
 \eta_1)...\hat \rho(\eta_\ell) 
\delta\left(g-\left[\prod_{i=1}^\ell (1+\eta_i)-1\right]\right) \ .
 \label{def_Omega}
\end{equation}

The equation for $ \rho (\eta_0)$ reads:

\begin{eqnarray}
 \rho(\eta_0) &=& \frac{1}{1-t} \sum_{k=0}^\infty \sum_{m=1}^{k}
\sum_{n=0}^{k-m} C_{k,m,n} \sum_{k'=0}^\infty \sum_{m'=1}^{k'}
\sum_{n'=0}^{k'-m'} C_{k',m',n'}  \int dg_+ dg_-  dg_+' dg_-' \times  \nonumber
\\ & \times & \Omega^{(m)}(g_+) \Omega^{(n)}(g_-)
 \Omega^{(m')}(g_+')  \Omega^{(n')}(g_-') \times \nonumber
\\ & \times &
\delta\left(\eta_0- \frac{g_+ g_+'}{ g_+ (1+g_-')+g_+' (1+g_-)+(1+g_-)(1+g_-')}\right)
\label{EQ_RHO_HAT}
\end{eqnarray}

Clearly, the function $\Omega^{(\ell)}(g)$ can be seen as the $\ell$-th
convolution of a certain function after an appropriate change of variables.
One is lead to introduce the variables $\phi_i$ and $x$ defined by:
\begin{equation}
\phi_i= \log(1+\eta_i) \ \ \ , \ \ \ x=\log(1+g) \ ,
\end{equation} 
and we call $S(\phi_i)$ , $T^{(m)}(x)$ their probability distributions. 
Equation(\ref{def_Omega}) shows that:
\begin{equation}
T^{(m)}(x)=\int d\phi_1...d\phi_m  S(\phi_1)...S(\phi_m)
\delta\left(x-\sum_{i=1}^m \phi_i\right) \; \;.
\end{equation}
where $\phi_i \in [0,\ln 2]$ and $x \in [0,m\ln 2]$.

In order to simplify the self-consistency equation, we introduce the
joint probability distribution
\begin{equation}
R(x_+,x_-) \equiv \frac{1}{\sqrt{1-t}} \sum_k 
\sum_{m= 1}^k  \sum_{n=0}^{k-m} C_{k,m,n} T^{(m)}(x_+) T^{(n)}(x_-)
\label{Rdef}
\end{equation}
which is normalized: $\int dx_+ dx_- R(x_+,x_-) =1$.

We may show that $R$ factorizes by introducing its Fourier Transform:
Using the coefficients (\ref{Ckmn_def}), the triple series can be resummed
and expressed in terms of the Fourier transform $\hat S(q)$ of $S(x)$: 
\begin{eqnarray}
&\int& dq_+ dq_- R(x_+,x_-) e^{i (q_+ x_+ + q_- x_-)} =\nonumber \\
&=& \frac{e^{-3 \alpha}}{\sqrt{1-t}} \sum_k \frac{(3 \alpha)^k}{k!}
\left( 
   \left[t+\frac{1-t}{2} \hat S(q_+)+\frac{1-t}{2}\hat S(q_-)\right]^k
  -\left[t+\frac{1-t}{2} \hat S(q_-) \right]^k 
\right) \nonumber \\
&=& \frac{1}{\sqrt{1-t}} 
\exp\left[3 \alpha 
  \left(t-1+ \frac{1-t}{2} \hat S(q_+)+\frac{1-t}{2}\hat S(q_-)\right)
\right]\nonumber \\
& & -\frac{1}{\sqrt{1-t}}
\exp\left[3 \alpha \left(t-1+\frac{1-t}{2} \hat S(q_-) \right) \right] \ .
\end{eqnarray}

Rearranging the above expression and taking the inverse transformation we
find for $R$
\begin{equation}
R(x_+,x_-)= A(x_+) B(x_-) \ ,
\label{ReqAB}
\end{equation}
where
\begin{eqnarray}
A(x_+) &\equiv& \frac{1}{e^{3 \alpha (1-t)/2}-1} \int \frac{dq_+}{2
\pi}  e^{-i q_+ x_+} \left( e^{\frac{3\alpha}{2}(1-t) \hat
S(q_+)}-1 \right) \nonumber \\
B(x_-)&\equiv& \frac{1}{e^{3 \alpha (1-t)/2}}  \int \frac{dq_-}{2
\pi}  e^{-i q_- x_-} 
 e^{\frac{3\alpha}{2}(1-t) \hat
S(q_+)}
\end{eqnarray}

We may now write the self-consistency equation in a tractable form.
Defining the variable $\phi_0$ associated with $\eta_0$ as:
\begin{equation}
\eta_0=
\frac{(e^{x_+}-1)(e^{x_+'}-1)}{(e^{x_+}-1) e^{x_-'}+ (e^{x_+'}-1) e^{x_-} + e^{x_-} e^{x_-'}}
=e^{\phi_0}-1
\end{equation}
we transform the equation for $\hat \rho(\eta_0)$ into an equation for
$S(\phi_0)$
\begin{eqnarray}
S(\phi_0)&=&\int dx_+ dx_- dx_+'dx_-' R(x_+,x_-) R(x_+',x_-')
\nonumber \\
&& \delta\left(\phi_0-\log\left[1+ \frac{(e^{x_+}-1)(e^{x_+'}-1)}
{(e^{x_+}-1) e^{x_-'}+(e^{x_+'}-1) e^{x_-}+e^{x_-}e^{x_-'}}\right]\right)
\label{EQ_SPHI_1}
\end{eqnarray}
This is the equation that we have used in order to solve the problem numerically.

Let us mention however that 
a  series of further simplifications may also be written.
It is easy to verify that 
\begin{equation}
S(\phi_0)=\int dz dz' C(z) C(z') \delta\left(\phi_0-\log\left[1+\frac{1}{z+z'+z z'}\right]\right)
\label{EQ_SPHI_2}
\end{equation}
where
\begin{equation}
C(z)\equiv \int  dx_+ A(x_+) dx_- B(x_-) \delta\left(z-\frac{e^{x_-}}{e^{x_+}-1}\right)
\end{equation}
Moreover, if we perform in the above the change of variable $1+z=e^\zeta$,
defining the  distribution 
\begin{equation}
D(\zeta)=C(z) \frac{d}{d\zeta} \ ,
\end{equation}
 we find:
\begin{equation}
S(\phi_0)=\int d\eta E(\eta)\delta\left(\phi_0-\log[\frac{e^\eta}{e^{\eta}-1}]\right)
\label{EQ_SPHI_3}
\end{equation}
where  $E(\eta)$ is the convolution of $D$ with itself:
\begin{equation}
 E(\eta) \equiv\int d\zeta d\zeta' D(\zeta)  D(\zeta')\delta(\zeta+\zeta'-\eta) \ .
\end{equation}

The transformations which we have written from $S \to A, B \to C \to D
\to E \to S$ can all be done using one dimensional integrals, Fourier
transforms, and changes of variables.  Hence this provides an
iterative mapping from the function $S(\phi)$ onto itself which can be
handled efficiently numerically. Which form of the self-consistency
equation to use is a matter of computational convenience.  It turns
out that, for our purpose, enough precision could be obtained from
(\ref{EQ_SPHI_2}) and we did not try to develop this alternative
computation.

We now proceed with the computation of the energy $\Phi(y)$ defined in
(\ref{PHIrsb}).  We shall be interested in evaluating it at large $y$,
using the solution for the distribution of u-surveys that we have just
found in this limit.

We start with the piece $\Phi_1(y)$. It consists of two terms:
\begin{equation}
\Phi_1(y)=\frac{3 \alpha}{y} \overline {\log C_0}-\frac{1}{y} \overline {\log A_0}
 \ ,
\label{phi1decomp}
\end{equation}
where the overline denotes the average over the population dynamics of
sect. \ref{sect_popdyn} and  where $C_0$ and $A_0$ are given by
\begin{eqnarray}
A_0&=&\int \prod_{\ell=1}^k \left[du_\ell 
Q_{\ell }(u_\ell) \right] \; 
 \exp\left[{y \vert \sum_{\ell=1}^k u_\ell \vert}\right] \ , \\
\frac{1}{C_0}&=&
  \int  dg dh P_1(g) P_2(h) 
  \exp \left[ y \hat w_{\bf J}(g,h)\right]\ .
\label{A0C0def}
\end{eqnarray}

The term $A_0$ is easily written in terms of the variables $g_+,g_-,g_0$:
\begin{equation}
A_0=\sum_q f_q^{(m,n)}({\bf \eta}) e^{y\vert q\vert}=
g_+({\bf \eta})+g_-({\bf \eta})+g_0({\bf \eta})
\end{equation}
Averaging over the population dynamics, we get for large $y$:
\begin{eqnarray}
\frac{1}{y}\overline{\log A_0}&=&\frac{1}{y}\sum_{k,m,n} C_{k,m,n}
\int dg_0 dg_+ dg_- P^{(m,n)}(g_+,g_0,g_-) \log[g_0 + g_+ + g_-]
\nonumber \\
&=&\frac{1}{y}\sum_{k,m,n} C_{k,m,n}
\int dx dz T^{(m)}(x) T^{(n)}(z) \log\left(e^x+e^z-1\right) \ .
\label{afirst}
\end{eqnarray}
Treating separately the $m\ge 1$ piece which can be resummed
as in (\ref{Rdef},\ref{ReqAB}), and the  $m=0$ piece, the sum over $k,m,n$ gives
\begin{equation}
\sum_{k,m,n} C_{k,m,n} T^{(m)}(x) T^{(n)}(z)=
\sqrt{1-t} A(x) B(z)+\delta(x) \sum_{k,n} C_{k,0,n} T^{(n)}(z)= B(x) B(z)  \ .
\label{UTILE}
\end{equation}
Putting this expression back into (\ref{afirst}),
we finally find
\begin{equation}
\frac{1}{y}\overline{\log A_0}= \frac{1}{y}
  \int dx dz B(x) B(z) \log\left(e^x+e^z-1\right) \ .
\label{A0res}
\end{equation}

We now turn to the second contribution, $C_0$, to (\ref{A0C0def});
before averaging over the iteration of the population dynamics, we have:
\begin{equation}
\frac{1}{C_0}=
\int  \prod_{i=1}^k \left[du_i Q_{i}(u_i)\right]
\prod_{j=1}^{k'} \left[dv_j Q_{k+j}(v_j)\right]
\exp\left[y \hat w_{\bf J}\left(\sum_{i=1}^{k} u_i,\sum_{j=1}^{k'} v_j\right)\right]
\label{gri}
\end{equation}
 Without loss of generality we assume
$J_1=J_2=1$;  we use the solution (\ref{Ansatz_3sat_2}), and denote
as before by $m,n,m',n'$ the numbers of various u-surveys appearing
in(\ref{gri}):
\begin{eqnarray}
k \; \; \; &\to& \text{ $m$ type $b_+$,  $n$ type $b_-$ and
$k-m-n$ type $a$} \nonumber \\
k' \; \; \; &\to& \text{ $m'$ type $b_+$,  $n'$ type $b_-$ and
$k'-m'-n'$ type $a$} \nonumber 
\end{eqnarray}
$C_0$ can then be written in terms of the $g$ variables as:
\begin{eqnarray}
\frac{1}{C_0}&=&\int dg_0 dg_+ dg_- P^{(m,n)}(g_+,g_0,g_-) \;
 dg_0' dg_+' dg_-' P^{(m',n')}(g_+',g_0',g_-') \nonumber \\
&&
\left(g_+ g_+' e^{-y}+ g_+[ g_-'+g_0']+
 g_+'[ g_- +g_0]+ [ g_-'+g_0'] [ g_-+g_0] \right)
\end{eqnarray}
For $y$ large we can drop the term in $\exp(-y)$; averaging over the iteration
of the population, we get
\begin{eqnarray}
- \overline{\log C_0}&=& \sum_{k,m,n}
\sum_{k',m',n'} C_{k,m,n}  C_{k',m',n'}
\int dx dz \; T^{(m)}(x) T^{(n)}(z)\nonumber \\
&&\int dx' dz' \; T^{(m')}(x') T^{(n')}(z')
\log\left[(e^x-1)e^{z'}+(e^{x'}-1)e^{z}+e^{z}e^{z'}\right]\nonumber \\
&=&\int dx dz dx' dz' \; B(x) B(z) B(x') B(z')
\log\left[(e^x-1)e^{z'}+(e^{x'}-1)e^{z}+e^{z}e^{z'}\right]
\label{C0res}
\end{eqnarray}

Finally we now compute the $\Phi_2(y)$ piece given in (\ref{PHIrsb}).
 For a generic clause with coupling $J_1,J_2,J_3$, involving
the h-surveys $P_1(h_1)$, $P_2(h_2)$, $P_3(h_3)$, we have
\begin{eqnarray}
\exp[-y \Phi_2(y)] &=& \int dh_{1} dh_{2} dh_{3} P_{1}(h_1) P_{2}(h_2)
P_{3}(h_3) e^{-2 y \theta(J_1 h_1) \theta(J_2 h_2) \theta(J_3 h_3) \ .
}
\label{phi2deb}
\end{eqnarray}
We write as before $P_1(h_1)=\int \prod_{\ell=1}^{k_1} [du_\ell
Q_l(u_\ell)]$, and suppose that this set of $k_1$ u-surveys contains
$m_1$ u-surveys of type $b_+$, $n_1$ of type $b_-$ and $k_1-m_1-n_1$
of type $a$, characterized by the weights ${\bf \eta_1}$.  The same
decomposition is done for $P_2(h_2)$ (resp $P_3(h_3)$), where the
numbers of u-surveys of various types are $m_2,n_2,k_2-m_2-n_2$ and
the weights are ${\bf \eta_2}$(resp.  $m_3,n_3,k_3-m_3-n_3$,${\bf
\eta_3}$).  For this iteration, the expression (\ref{phi2deb}) for
$\Phi_2(y)$ can be reexpressed as:
\begin{eqnarray}
\exp[-y \Phi_2(y)] &=& \sum_{q_1,q_2,q_3} f_{q_1}^{(m_1,n_1)}({\bf \eta_1})
 f_{q_2}^{(m_2,n_2)}({\bf \eta_2}) f_{q_3}^{(m_3,n_3)}({\bf \eta_3})
\exp[-2 y \theta(q_1)\theta(q_2)\theta(q_3)] \nonumber \\
&\simeq_{y \to \infty}& \prod_{i=1}^3 \left[ g_+\left({\bf \eta_i}\right)+
g_-\left({\bf \eta_i}\right)+g_0\left({\bf \eta_i}\right)\right]-
\prod_{i=1}^3 \left[ g_+\left({\bf \eta_i}\right)\right]
\end{eqnarray}
The average over the population gives:
\begin{eqnarray}
\overline {\Phi_2(y)} &=& -\frac{1}{y}
\sum_{k_1,m_1,n_1}\; \sum_{k_2,m_2,n_2}\; \sum_{k_3,m_3,n_3}
 C_{k_1,m_1,n_1} \; C_{k_2,m_2,n_2}\; C_{k_3,m_3,n_3}  \nonumber \\
&\times & \prod_{i=1}^3 \left\{ \int dx_i dz_i T^{(m_i)}(x_i)
T^{(n_i)}(y_i) \log \left[ \prod_{i=1}^3 (e^{x_i}+e^{z_i}-1)-
\prod_{i=1}^3 (e^{x_i}-1) \right]\right\}\nonumber \\
&=& -
 \frac{1}{y} \int \prod_{i=1}^3 dx_i dz_i B(x_i) B(z_i) 
\log \left[ \prod_{i=1}^3 (e^{x_i}+e^{z_i}-1)-\prod_{i=1}^3 (e^{x_i}-1) \right]
\label{phi2res}
\end{eqnarray}
Grouping together the  contributions (\ref{phi1decomp},\ref{A0res},\ref{C0res},\ref{phi2res}),
we find the total zero temperature free energy density
$
\Phi(y)=\Phi_1(y)-2 \alpha \Phi_2(y)
$ 
given in (\ref{Phifinal}).

\section{Free energy for one given sample}

\begin{figure}
\centerline{
        \epsfxsize=12.0cm
        \epsfbox{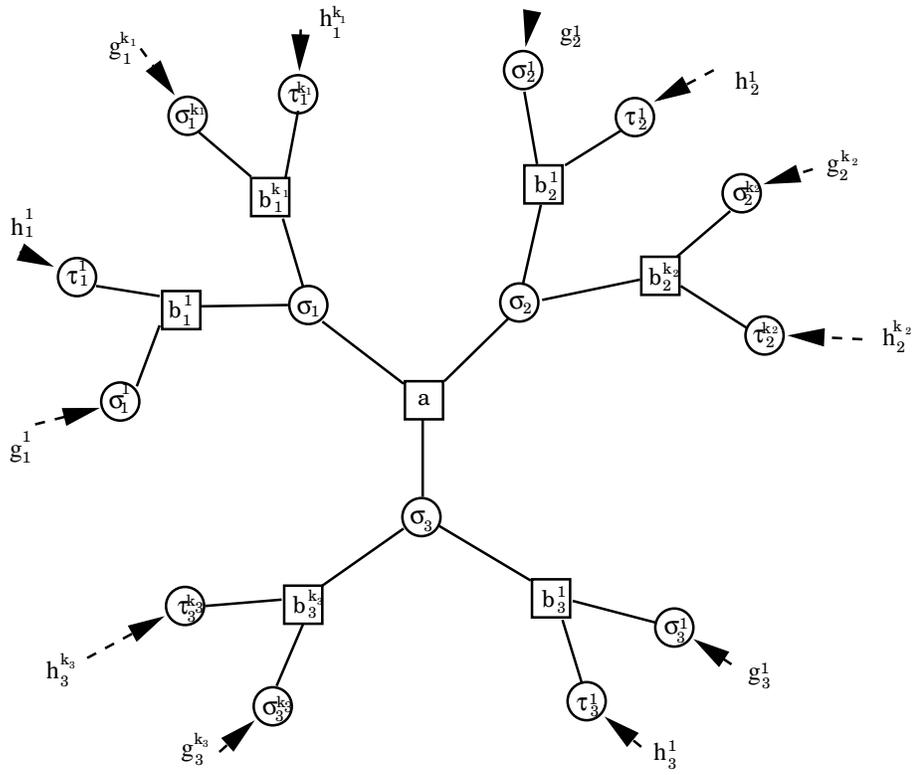}
}
  \caption{ The set of nodes with which one computes the free energy shift 
$\Phi^f_{a}(y)$.
   }
\label{fig_app}
\end{figure}

\begin{figure}
\centerline{
        \epsfxsize=12.0cm
        \epsfbox{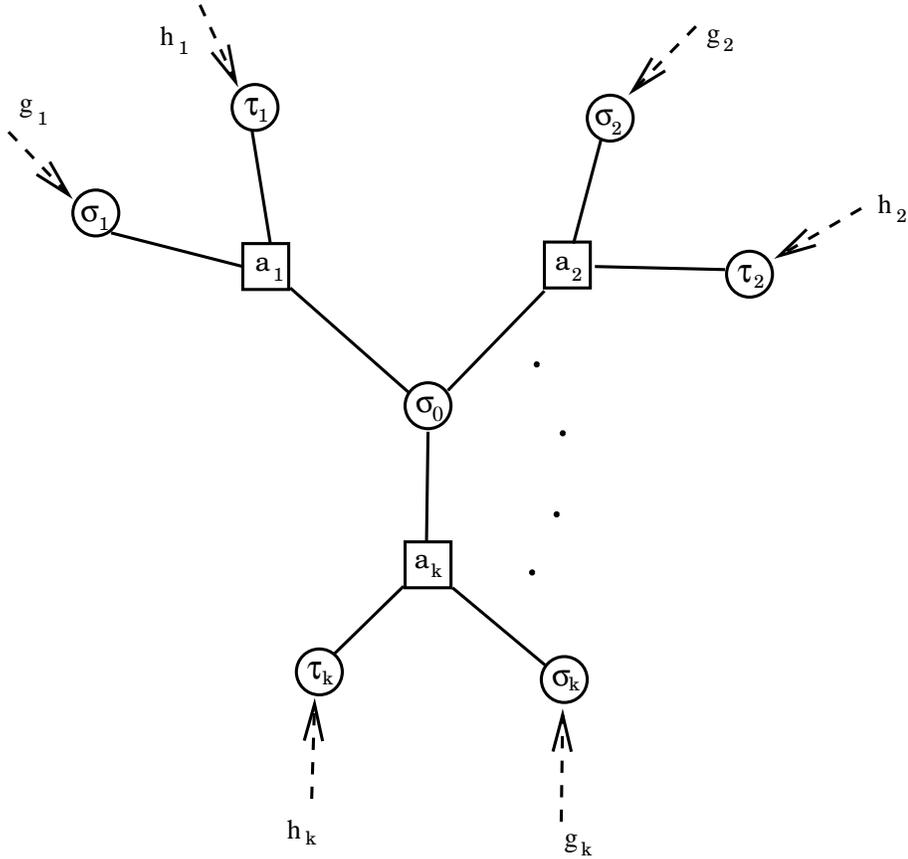}
}
  \caption{ The set of nodes with which one computes the free energy shift 
$\Phi^v_{0}(y)$.
   }
\label{fig_app2}
\end{figure}

Let us explain here how to compute the zero temperature free energy
for one given sample.  We start from the contribution of one given
factor node $a$.  We shall look at a somewhat large part of the graph
containing $a$ (see fig.\ref{fig_app}).  We call $\sigma_1,
\sigma_2,\sigma_3$ the three spins connected to it. One of these
spins, $\sigma_r$ is connected, beside $a$, to $k_r$ other function
nodes which we call $b_r^1,...,b_r^{k_r}$.  The function node $b_r^s$
is connected, beside $\sigma_r$, to two other spins which we call
$\sigma_r^s$ and $\tau_r^s$, and the cavity fields onto them are
called $g_r^s$ and $h_r^s$ (see fig.\ref{fig_app}).  In the absence of
the spins $\sigma_1, \sigma_2,\sigma_3$ and of all the function nodes
$b_r^s$, the ground state energy of the system would be
\begin{equation}
E_{init}=-\sum_{r=1}^3 \sum_{s=1}^{k_r} \left( \vert g_r^s\vert+\vert  h_r^s\vert \right)
\end{equation}
Adding the spins $\sigma_1,
 \sigma_2,\sigma_3$ and  all the function nodes $b_r^s$,
the ground state energy becomes
\begin{eqnarray}
E_{fin}&=&\min_{\sigma_1,\sigma_2,\sigma_3}\left\{E_a(\sigma_1,\sigma_2,\sigma_3)+
  \sum_{r=1}^3 \sum_{s=1}^{k_r}
 \left( \min_{\sigma_r^s,\tau_r^s} \left[ E_{b_r^s}(\sigma_r,\sigma_r^s,\tau_r^s)-
g_r^s \sigma_r^s -h_r^s \tau_r^s\right]\right)\right\} \nonumber \\ \nonumber
&=&\min_{\sigma_1,\sigma_2,\sigma_3}\left\{E_a(\sigma_1,\sigma_2,\sigma_3)
-\sum_{r=1}^3\sigma_r \sum_{s=1}^{k_r} \hat u_{\bf J}(g_r^s,h_r^s)\right\}\\
&&-\sum_{r=1}^3 \sum_{s=1}^{k_r}\left(\hat w_{\bf J}(g_r^s,h_r^s)
-\vert g_r^s\vert-\vert  h_r^s\vert\right)
\end{eqnarray}
The zero temperature free energy shift induced by the addition of
all these nodes is given by:
\begin{equation}
e^{-y \Phi^f_{a}(y)}=
\int e^{-y(E_{fin}-E_{init})}
\end{equation}
where the integral is over all the $g_r^s,h_r^s$ fields, each with a
probability distribution given by its h-survey.
We can use the iteration equation (\ref{Q1siter}) in order to simplify this complicated
integral through the use of u-surveys:
\begin{equation}
e^{-y \Phi^f_{a}(y)}=
\int \prod_{r=1}^3\prod_{s=1}^{k_r}
\left[\frac{1}{C_{b_r^s \to r}} du_r^s Q_{b_r^s \to r}(u_r^s)\right]
\exp\left(-y\min_{\sigma_1,\sigma_2,\sigma_3}\left\{E_a(\sigma_1,\sigma_2,\sigma_3)
-\sum_{r=1}^3\sigma_r \sum_{s=1}^{k_r}  u_r^s\right\}\right)
\label{appbond}
\end{equation}
We now compute the contribution from one given variable node  $i=0$. We 
use the notations of fig.\ref{fig_app2}, calling $a_r$ the $k$ function 
nodes to which it is connected ($r \in \{ 1,..,k\}$). The function 
node $a_r$ is connected, beside $\sigma_0$, to the spins 
$\sigma_r,\tau_r$, and we call 
$g_r$ (resp. $h_r$) the cavity field on $\sigma_r$ (resp. $\tau_r$). In 
the absence of spin $\sigma_0$ and of the function nodes connected to it, 
the
ground state energy of the system would be:
\begin{equation}
E_{init}= -\sum_{r=1}^k \left(\vert g_r\vert + \vert h_r \vert\right)
\end{equation}
Adding the new spin and the function nodes $a_r$, the ground state energy 
becomes:
\begin{eqnarray}
E_{fin}&=& \min_{\sigma_0}\sum_{r=1}^k\left(\min_{\sigma_r,\tau_r} \left[
E_{a_r}(\sigma_0,\sigma_r,\tau_r)-g_r\sigma_r-h_r
\tau_r\right]\right) \nonumber \\
&&
=\min_{\sigma_0}\left(-\sigma_0 \sum_{r=1}^k\hat u_{\bf J_r}(g_r,h_r)\right)
-\sum_{r=1}^k\hat w_{\bf J_r}(g_r,h_r)=-\vert \sum_{r=1}^k\hat u_{\bf 
J_r}(g_r,h_r) \vert -\sum_{r=1}^k\hat w_{\bf J_r}(g_r,h_r)
\end{eqnarray}
The zero temperature free energy shift induced by the addition of these 
nodes is given by:
\begin{equation}
e^{-y\Phi_0^v(y)}=\int e^{-y(E_{fin}-E_{init})}
\end{equation}
where the integral is over all the $g_r,h_r$ fields, each with a 
probability distribution given by its h-survey. As usual, this can be 
simplified by the use of u-surveys derived from the iteration eq. 
(\ref{Q1siter}):
\begin{equation}
e^{-y\Phi_0^v(y)}=\int \prod_{r=1}^k \left[ \frac{du_r}{C_{{a_r}\to 0}} \; 
Q_{{a_r}\to 0} (u_r)\right] e^{y \vert \sum u_r \vert}
\label{appsite}
\end{equation}
A little thought shows that, when computing the total zero
temperature free energy $\Phi(y)=\sum_a \Phi_a^f(y)-\sum_i(n_i-1)
\Phi_i^v(y)$, one is correctly counting each node once. In particular,
in the limit of $y \to 0$, $\Phi(y)$ reduces to the sum of the energy
of all factor nodes, as it should.  The same reasoning shows that the
"$1/C$" factors in (\ref{appbond}) and (\ref{appsite}) actually
cancel, so that one can forget these normalisations for the
computation of $\Phi(y)$, as was done in the text in formulae
(\ref{free_onesamp1},\ref{free_onesamp2}).


\begin{thebibliography}{1}
\bibitem{fields_comment} Note that the existence of finite local
fields is true for states of arbitrary energy. In particular,
considering the zero energy states, selected by a reweighting $y \to
\infty$, one sees from (\ref{Ansatz_3sat_2}) that their non trivial
cavity fields scale as $e^{-y}$; however when computing the local
fields as in (\ref{local_field}), the extra reweighting exactly
compensates this scaling, giving finite local fields also when $y \to
\infty$.
\end{thebibliography}
\end{document}